\documentclass[longauth]{aa}
\newcommand{\degree}{$^{\mathrm{o}}$}

\newcommand{\micron}{$\rm \mu m$}

\newcommand{\Msun}{$\rm M_{\odot}$}
\newcommand{\Lsun}{$\rm L_{\odot}$}

\graphicspath{{./}{figs/}}
\usepackage[]{hyperref}

\hypersetup{breaklinks=true,colorlinks=true,urlcolor=blue,linkcolor=blue,citecolor=blue,pagecolor=red,bookmarks=true}

\usepackage{graphicx}
\usepackage{txfonts}

\begin{document}

   \title{Gap, shadows, spirals, and streamers: SPHERE observations of binary-disk interactions in \object{GG\,Tau\,A}\thanks{Based on observations performed with VLT/SPHERE under program ID 198.C-0209(N)}}
   \titlerunning{SPHERE observations of GG\,Tau\,A}
   \authorrunning{M. Keppler et al.}

   \author{
      M.~Keppler\inst{1}
    \and A.~Penzlin\inst{2}
    \and M.~Benisty\inst{3,4}
    \and R.~van\,Boekel\inst{1}
    \and T.~Henning\inst{1}
    \and R.~G.~van\,Holstein\inst{5,6}
    \and W.~Kley\inst{2}
    \and A.~Garufi\inst{7}
    \and C.~Ginski\inst{8,5}
    \and W.~Brandner\inst{1}
    \and G.~H.-M.~Bertrang\inst{1}
    \and A.~Boccaletti\inst{9}
    \and J.~de\,Boer\inst{5}
    \and M.~Bonavita\inst{10,11}
    \and S.~Brown\,Sevilla\inst{1}
    \and G.~Chauvin\inst{3,4}
    \and C.~Dominik\inst{8}
    \and M.~Janson\inst{12}
    \and M.~Langlois\inst{13,14}
    \and G.~Lodato\inst{15}
    \and A.-L.~Maire\inst{16,1} %
    \and F.~M\'{e}nard\inst{3}
    \and E.~Pantin\inst{17} %
    \and C.~Pinte\inst{18,3} %
    \and T.~Stolker\inst{19}
    \and J.~Szul\'{a}gyi\inst{20}
    \and P.~Thebault\inst{21}
    \and M.~Villenave\inst{3}
    \and A.~Zurlo\inst{22,23,14} %
    \and P.~Rabou\inst{3}
    \and P.~Feautrier\inst{3}
    \and M.~Feldt\inst{1}
    \and F.~Madec\inst{14}
    \and F.~Wildi\inst{24}
    }
    \institute{
    Max Planck Institute for Astronomy, K\"{o}nigstuhl 17, 69117, Heidelberg, Germany \email{keppler@mpia.de}
    \and Institut f\"{u}r Astronomie und Astrophysik, Universit\"{a}t T\"{u}bingen, Auf der Morgenstelle 10, 72076 T\"{u}bingen, Germany
    \and Univ. Grenoble Alpes, CNRS, IPAG, 38000 Grenoble, France
    \and Unidad Mixta Internacional Franco-Chilena de Astronom\'{i}a (CNRS, UMI 3386), Departamento de Astronom\'{i}a, Universidad de Chile, Camino El Observatorio 1515, Las Condes, Santiago, Chile
    \and Leiden  Observatory, Leiden University, P.O.  Box 9513, 2300 RA Leiden, The Netherlands
    \and European Southern Observatory, Alonso de C\'{o}rdova 3107, Casilla 19001, Vitacura, Santiago, Chile
    \and INAF, Osservatorio Astrofisico di Arcetri, Largo Enrico Fermi 5, 50125 Firenze, Italy
    \and Anton Pannekoek Institute for Astronomy, Science Park 904, NL-1098 XH Amsterdam, the Netherlands
    \and LESIA, CNRS, Observatoire de Paris, Universit\'{e} Paris Diderot, UPMC, 5 place J. Janssen, 92190 Meudon, France
    \and SUPA, Institute for Astronomy, University of Edinburgh, Blackford Hill, Edinburgh EH9 3HJ, UK
    \and Centre for Exoplanet Science, University of Edinburgh, Edinburgh EH9 3HJ, UK
    \and Department of Astronomy, Stockholm University, AlbaNova University Center, 106 91 Stockholm, Sweden
    \and CRAL,  UMR  5574,  CNRS,  Universit\'{e} de Lyon, Ecole Normale Sup\'{e}rieure de Lyon, 46 all\'{e}e d'Italie, 69364 Lyon Cedex 07, France
    \and Aix Marseille Univ., CNRS, CNES, LAM, Marseille, France
    \and Dipartimento di Fisica, Universit\`{a} degli Studi di Milano, Via Giovanni Celoria 16, I-20133 Milano, Italy
    \and STAR Institute, Universit\'{e} de Li\`{e}ge, All\'{e}e du Six Ao\^{u}t 19c, 4000 Li\`{e}ge, Belgium
    \and AIM, CEA, CNRS, Universit\'{e} Paris-Saclay, Universit\'{e} Paris Diderot, Sorbonne Paris Cit\'{e}, Gif-sur-Yvette, France
    \and School of Physics and Astronomy, Monash University, Clayton, VIC 3168, Australia
    \and Institute for Particle Physics and Astrophysics, ETH Zurich, Wolfgang-Pauli-Strasse 27, 8093 Zurich, Switzerland
    \and Institute for Theoretical Astrophysics and Cosmology, Institute for Computational Science, University of Z\"{u}rich, Winterthurerstrasse 190, CH-8057 Z\"{u}rich, Switzerland
    \and LESIA-Observatoire de Paris, UPMC Univ. Paris 06, Univ. Paris-Diderot, France
    \and N\`{u}cleo de Astronom\'{i}a, Facultad de Ingenier\'{i}a y Ciencias, Universidad Diego Portales, Av. Ejercito 441, Santiago, Chile
    \and Escuela de Ingenier\'ia Industrial, Facultad de Ingenier\'ia y Ciencias, Universidad Diego Portales, Av. Ejercito 441, Santiago, Chile
    \and Geneva Observatory, University of Geneva, Chemin des Mailettes 51, 1290 Versoix, Switzerland
    }
   \date{Received ---; accepted ---}

  \abstract
    {
    A large portion of stars is found to be part of binary or higher-order multiple systems. The ubiquity of planets found around single stars raises the question of whether and how planets in binary systems form.
    Protoplanetary disks are the birthplaces of planets, and characterizing them is crucial in order to understand the planet formation process. 
    }
    {Our goal is to characterize the morphology of the GG\,Tau\,A disk, one of the largest and most massive circumbinary disks. We also aim to trace evidence for binary-disk interactions. }
    {We obtained observations in polarized scattered light of GG\,Tau\,A using the SPHERE/IRDIS instrument in the H-band filter. We analyzed the observed disk morphology and substructures. We ran 2D hydrodynamical models to simulate the evolution of the circumbinary ring over the lifetime of the disk. }
    {The disk and also the cavity and the inner region are highly structured, with several shadowed regions, spiral structures, and streamer-like filaments. Some of these are detected here for the first time. The streamer-like filaments appear to connect the outer ring with the northern arc. Their azimuthal spacing suggests that they may be generated through periodic perturbations by the binary, which tear off material from the inner edge of the outer disk once during each orbit.
    By comparing observations to hydrodynamical simulations, we find that the main features, in particular, the gap size, but also the spiral and streamer filaments, can be qualitatively explained by the gravitational interactions of a binary with a semimajor axis of $\sim$35\,au on an orbit coplanar with the circumbinary ring.}
  {}
   \maketitle
%
%-------------------------------------------------------------------
%-------------------------------------------------------------------

\section{Introduction} \label{sec:intro}
Almost half of all main-sequence solar-type stars are found in binary or higher-order multiple systems \citep[e.g., ][]{Raghavan+10,Duchene+Kraus13}, and it is thought that the fraction of multiple systems is even higher among pre-main sequence stars \citep[e.g., ][]{Duchene+99, Kraus+11}. More than 4000 detections of extrasolar planets around single stars to date show that the assembly of planetary bodies is a common byproduct of star formation. The high abundance of multiple stars on the one hand and planetary companions on the other hand thus raises the question about the possible formation pathways and prevalence of planets in multiple systems.

While our understanding of the building-up of planets within protoplanetary disks around single stars has significantly advanced in the past years, less is known about the conditions of planet formation in multiple systems \citep[e.g.,][]{Thebault+15}. 
In contrast to the single-star case, the evolution of material in the circumbinary and individual circumstellar disks in multiple systems will (depending on the binary parameters such as mass ratio, orbital separation, and eccentricity) be dominated by the gravitational perturbation of the central binary. As a consequence, the binary-disk interaction has severe implications for the planet formation process. Tidal interactions exerted by the binary are expected to truncate the individual circumstellar disks, reducing their masses, outer radii, and viscous timescales \citep[e.g.,][]{Papaloizou+77,Artymowicz+94,Rosotti+Clarke18}. In addition, the tidal torques will truncate the circumbinary disk from the inner edge by opening a large inner cavity. Despite the resulting separation of circumbinary and circumstellar material, gas streams through the gap may form, supplying the circumstellar disks with material from the outer circumbinary disk \citep[e.g.,][]{Artymowicz+96,Munoz+19_accretion}. While observational trends infer binary interaction to be indeed destructive for disks in many cases \citep[e.g.,][]{Bouwman+06,Duchene+10,Harris+12,Cox+17,Akeson+19,Manara+19}, potentially impeding the formation of planets, several massive disks around binary systems are known and have been observed at high angular resolution \citep[e.g., UY\,Aur, HD142527, HD\,34700\,A; ][]{Hioki+07,Tang+14,Avenhaus+17,Monnier+19}.

Despite the potential complications for planet formation induced by the gravitational perturbations from the binary, more than 100 planets in binary systems have already been discovered \citep[e.g.,][]{Martin+18,Bonavita+20}\footnote{see also \url{http://www.univie.ac.at/adg/schwarz/multiple.html} \citep{Schwarz+16}}. Most of these planets are found to orbit only one of the binary stars (i.e., `S-type', i.e., circumstellar planets). The reason for this certainly is that the radial velocity and transit photometry methods, which represent the most successful planet detection methods in terms of numbers, are strongly biased toward planets on short orbital periods. Nevertheless, about 20 planets have been discovered on orbits surrounding  both binary components (i.e., `P-type', i.e., circumbinary planets) \citep[e.g., ][]{Doyle+11,Orosz+19}. The
statistical analysis of the first direct-imaging survey dedicated to finding planets orbiting two stars suggests that the distributions of planets and brown dwarfs are indistinguishable between single and binary stars within the error bars \citep{Bonavita+16,Asensio+18}. This implies that planet formation in multiple systems, and in particular, in circumbinary disks indeed occurs.

Most of the circumbinary planets were detected with the Kepler space telescope on close ($\lesssim$\,1\,au) orbits around eclipsing binary systems. Interestingly, they seem to orbit their host systems close to the stability limit, implying that migration processes and planet-disk interactions may have played a crucial role during their early evolution \citep[e.g.,][]{Kley+14}. 
It is therefore clear that the observation and characterization of circumbinary disks provide the unique opportunity of testing the conditions and setup for possible planet formation in multiple systems.  

One of these cases is GG\,Tau. Located at a distance of 150\,pc \citep[see Sect. \ref{sect:stellar_parameters}; ][]{Gaia2016,Gaia2018}, GG\,Tau is a young \citep[$\sim$1-4 Myr; ][]{White+99, Hartigan+Kenyon03, Kraus+Hillenbrand09} hierarchical quintuple system composed of two main components, GG\,Tau\,Aa/b and GG\,Tau\,Ba/b, at a projected separation of about $\sim$\,10\arcsec \ ($\sim$\,1500\,au) \citep{Leinert91,Leinert93}. The northern and more massive binary, GG\,Tau\,Aa/b (projected separation $\sim$ 0.25\arcsec, corresponding to $\sim$\,38\,au) is surrounded by a bright and well-studied circumbinary disk. Recent interferometric observations suggest that the secondary component, GG\,Tau\,Ab, is a binary itself (GG\,Tau\,Ab1/2) at a projected separation of about 31.7\,mas ($\sim$4.8\,au) \citep{diFolco+14}.

The circumbinary disk around GG\,Tau\,A is observed as a large and massive disk with a cleared cavity. While the gaseous disk extends out to more than $\sim$\,850\,au and reveals a reduced amount of gas in the inner region \citep[e.g.,][]{Guilloteau+99,Dutrey+14,Phuong+20}, the population of large dust grains observed at (sub-)millimeter wavelengths is confined within a narrow ring surrounding a deeply depleted dust cavity, spanning a full width of $\sim$\,60-80\,au centered at a radial distance of about 250\,au with respect to the system barycenter \citep[e.g.,][]{Andrews+14,Dutrey+14,Tang+16}. Scattered-light observations in the optical, near- and thermal infrared regime infer that the inner edge of the outer disk of the small-grain population is located at about $\sim$190-200\,au \citep[e.g.,][]{Krist+02,Duchene+04,Itoh+14,Yang+17}. Such a radial concentration of dust is indicative of particles being trapped within a pressure maximum at the edge of the cavity, as expected for binary-disk interactions \citep[e.g.,][]{DeJuanOvelar+13,Cazzoletti+17}.

To what extent the tidal interactions of GG\,Tau\,Aa/b are responsible for the observed gap size has remained controversial, however. Because the radial location of the gas pressure maximum depends on the binary semimajor axis and eccentricity \citep[e.g.,][]{Artymowicz+94}, the knowledge of the binary orbit is required in order to compare the observed gap size with theoretical predictions. Based on almost two decades of orbital monitoring, a best-fit orbit with a semimajor axis of 36\,au and an eccentricity of 0.28 has been established \citep{Koehler+11}. However, this orbital solution assumes that the orbit is coplanar with the circumbinary ring; when this assumption is relaxed, the orbital solution is less well constrained and allows for larger orbit sizes. Several theoretical studies have concluded that in order to explain the observed gap size of $\sim$190\,au, the binary orbit should have a semimajor axis of about $\sim$\,65\,au, that is, about one-third of the gap size. To still remain consistent with the astrometric constraints, such a large binary orbit would have to be misaligned with respect to the circumbinary disk  \citep[e.g.,][]{Beust+Dutrey05,Cazzoletti+17,Aly+18}. It is clear that the respective geometry and orientation of binary orbit and circumstellar and circumbinary disk will have a severe effect on the potential of planet formation. Therefore, a detailed knowledge of these parameters is required.  

We present new high-resolution ($\sim$\,0.04\,\arcsec) near-infrared polarimetric observations of the GG\,Tau\,A system obtained with the SPHERE instrument. Our observations reveal the circumbinary environment at unprecedented detail. We confirm previously known disk substructures and reveal new features within the circumbinary disk. We compare our observations to hydrodynamical simulations in order to investigate whether the observed structures can be explained by binary-disk interactions. Our paper is structured as follows: first, we revise the stellar parameters of GG\,Tau\,A in Sect. \ref{sect:stellar_parameters}, followed by the presentation of our observations in Sects. \ref{sec:observations} \& \ref{sec:results}. Section \ref{sec:modeling} presents our modeling efforts, which are discussed in context with the observations in Sect. \ref{sec:discussion}. \\

\begin{table}[b]
\caption{Properties of the GG\,Tau\,A system assumed in this study.}
\label{table:stellar_properties}
\begin{tabular}{lcccc}
\hline \hline
\textbf{Stellar parameters}      & Aa       & Ab1     & Ab2   &  ref. \\
\hline 
Spectral type           & M0       & M2      & M3    &  a,b \\
L        [\Lsun]        & 0.44     & 0.153   & 0.077 &  a,b,c \\
Teff     [K]            & 3900     & 3400    & 3200  &  d \\
Mass     [\Msun]        & 0.65     & 0.30    & 0.20  &  e \\
Age      [Myr]          & 2.8      & 2.8     & 3.1   &  e \\
\hline
\textbf{Disk properties}     & \multicolumn{3}{c}{}                  &  \\
\hline
inclination         & \multicolumn{3}{c}{37$\pm$1\degree} & f \\
position angle      & \multicolumn{3}{c}{277$\pm$1\degree}& f \\
\hline
\hline
\end{tabular}
\tablefoot{\footnotesize{ a) \cite{Hartigan+Kenyon03}, b) \cite{diFolco+14}, c) \cite{Brauer+19}, d) \cite{Rajpurohit+13}, e) this work, f) \cite{Guilloteau+99}  }}
\end{table}

\begin{figure*}[bt]
    \begin{centering}
    \includegraphics[width=0.95\textwidth]{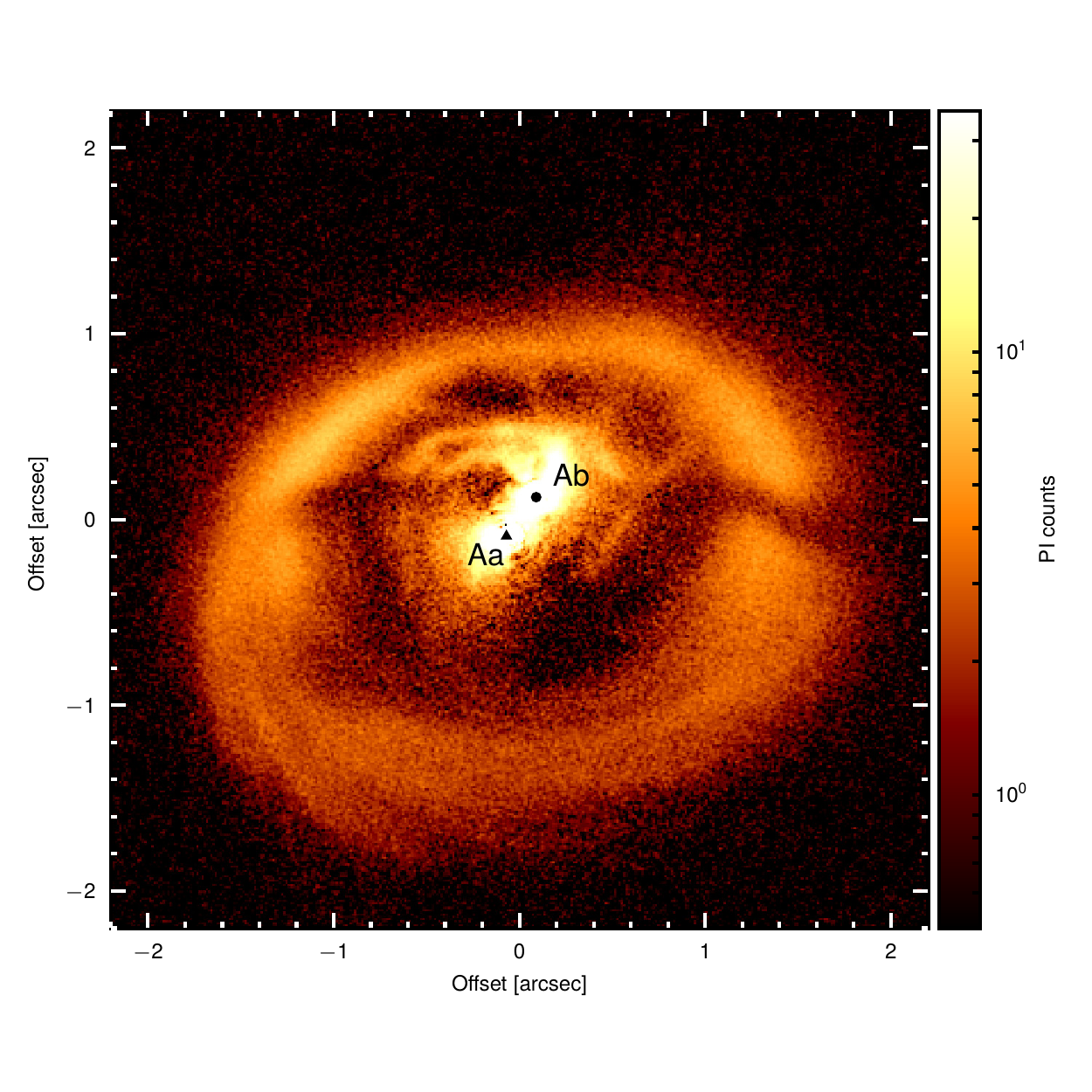}
    \end{centering}
   \caption{SPHERE polarized intensity ($PI$) image of GG\,Tau\,A. The image is centered on the expected location of the system's center of mass. The locations of GG\,Tau\,Aa and Ab are marked by a black triangle and circle, respectively. North is up and east is to the left.}  \label{fig:obs_PDI}
\end{figure*}

%--------------------------------------------------------------------
%--------------------------------------------------------------------

\section{Stellar properties}\label{sect:stellar_parameters}

Although several authors have studied the stellar properties of GG Tau A \citep[e.g., ][]{White+99, Hartigan+Kenyon03, Kraus+Hillenbrand09}, the discovery of the binarity of GG\,Tau\,Ab by \cite{diFolco+14} needs to be taken into account. In this work, we re-calculated the stellar masses and ages following this notion and the newly known distance $d$ to the system. A negative parallax has been reported for GG Tau A \citep{Gaia2018}, probably because of its binarity \citep[see also][]{Luhman+18}, with an absolute value of 6.65\,mas. GG\,Tau\,B, lying about 10\arcsec \ farther south than GG\,Tau\,A, has a positive parallax of 6.66\,mas. Because the two components are known to be bound, we used a parallax of 6.66\,mas (150\,pc) as a proxy for the distance of GG\,Tau\,A. We note, however, that the parallax measured for GG\,Tau\,B is likely affected by its own binarity as well, such that the distance of the system remains somewhat uncertain.

We assumed spectral types of M0, M2, and M3 for GG\,Tau\,Aa, Ab1, and Ab2 and an extinction of 0.3\,mag for Aa and 0.45\,mag for Ab1/2, as determined by \cite{Hartigan+Kenyon03} and \cite{diFolco+14}.
The corresponding stellar effective temperatures were obtained using the temperature scale of \cite{Rajpurohit+13} calibrated by their NTT spectra. We further assumed stellar luminosities derived by \cite{Hartigan+Kenyon03}, rescaled to 150\,pc, considering that their luminosity measured for Ab represents the sum of the luminosities of Ab1 and Ab2 with a respective luminosity ratio of $\sim$2:1 \citep[see][]{diFolco+14,Brauer+19}. 
We derived stellar masses and ages by comparing the locations of the GG\,Tau\,A components on a Hertzsprung-Russell diagram with those predicted by a set of five pre-main-sequence tracks \citep[Siess, PARSEC, MIST, Baraffe, Dartmouth; ][]{Siess+00,Bressan+12,Dotter+16,Choi+16,Baraffe+15,Dotter+08}. This yielded the following possible ranges for stellar masses and ages: 0.6-0.7\,\Msun \ and 2.4-3.1\,Myr for Aa, 0.3-0.5\,\Msun\ and 2.2-5.6\,Myr for Ab1, and 0.2-0.4\,\Msun\ and 2.7-10.0\,Myr for Ab2. We adopted the median of these values as our final stellar masses and ages: 0.65\,\Msun\ and 2.8\,Myr for Aa, 0.3 \Msun\ and 2.8\,Myr for Ab1, and 0.2 \Msun \ and 3.1\,Myr for Ab2.

Our inferred ages are well within the range of ages derived in previous studies \citep[$\sim$1-4\,Myr; ][]{White+99,Hartigan+Kenyon03,Kraus+Hillenbrand09}.  
Stars in multiple systems are generally assumed to form simultaneously and thus to be coeval. While Aa and Ab1 appear to be coeval according to our analysis, the age derived for Ab2 appears slightly older. However, increasing the luminosity of Ab2 by only 7\,\% reconciles the ages of all three stars. This has almost no effect on the derived mass of Ab2 because the evolutionary tracks run almost vertically in the Hertzsprung-Russell diagram at these young ages. 

While the median values of our inferred stellar masses add up to 1.15\,\Msun, which is slightly lower than the dynamical mass of the system derived through the CO observations of 1.37\,$\pm$\,0.08 \Msun \ \citep[][scaled to 150\,pc]{Guilloteau+99}, the range of possible stellar masses constrained by our models does not exclude a total mass of 1.37\,\Msun.
We note, however, that the determination of spectral types, effective temperatures, and luminosities, as well as the evolutionary models \citep[e.g., by not taking the effect of magnetic fields into account; ][]{Simon+19,Asensio+19} is hampered by some uncertainty, which might explain any discrepancy between our inferred values and those derived from the CO observations. Furthermore, our inferred total stellar mass might be underestimated if any of the components has an additional as yet undiscovered close-in stellar companion.

The circumbinary disk is observed at an inclination of 37\degree \ and at a position angle of 277\degree \ \citep{Guilloteau+99}. The system parameters are summarized in Table \ref{table:stellar_properties}.

%--------------------------------------------------------------------
%--------------------------------------------------------------------

\section{Observations and data reduction}\label{sec:observations}

GG\,Tau\,A was observed with SPHERE \citep{Beuzit+19} as part of the guaranteed-time observations (GTO) during the night of 2016 November 18. The IRDIS instrument \citep{Dohlen2008} was used in the dual-beam polarimetric imaging (DPI) mode \citep{Langlois2014,deBoer+20,vanHolstein+20}, applying the H-band filter (1.625\ \micron; pixel scale 12.25 mas/px), and the telescope operated in field-tracking mode.
One polarimetric cycle consisted of tuning the half-wave plate position at four different angles (0\degree, 45\degree, 22.5\degree \ , and 67.5\degree, respectively). At each of these positions, we took 15 frames with an exposure time of 4\,s each. A total of 11 polarimetric cycles was carried out, resulting in a total integration time on the science target of about 44 minutes. No coronagraph was used during the observations, inducing a slight saturation at the location of both Aa and Ab. Weather conditions were relatively stable during the observations (seeing at 500\,nm $\sim$\,0.6\,\arcsec-0.9\,\arcsec, coherence time $\sim$\,3\,ms, and wind speed $\sim$\,10\,m/s). We measured a point spread function (PSF) full width at half-maximum (FWHM) of about 43\,mas by fitting a Moffat pattern to the unsaturated images obtained with a neutral density filter.

The data were reduced using the IRDAP pipeline\footnote{\url{https://irdap.readthedocs.io}} \citep{vanHolstein+20}. In short, after basic steps of data reduction (dark subtraction, flat fielding, bad-pixel correction, and centering), the pipeline obtains the clean Stokes $Q$ and $U$ frames using the double-difference method. The data are then corrected for instrumental polarization and cross-talk effects by applying a detailed Mueller matrix model that takes the complete optical path of the light beam into account. 
After correcting for instrumental effects, the pipeline determines, and if desired, also subtracts, any remaining stellar polarization. This is measured by quantifying the flux in the Q and U images from regions without polarized disk emission. From the final $Q$ and $U$ images, a linear polarized intensity (\textit{PI}) image is then obtained, following \textit{PI}\,=\,$\sqrt{Q^2+U^2}$. This final image is corrected for true north \citep{Maire+16}. For details regarding the pipeline, we refer to \cite{vanHolstein+20}.
Finally, the images were recentered on the expected location of the center of mass, assuming a mass ratio between GG\,Tau\,Aa and GG\,Tau\,Ab1/2 of 0.77 (see Sect. \ref{sect:stellar_parameters}).

%--------------------------------------------------------------------
%--------------------------------------------------------------------

\section{Results}\label{sec:results}

The final \textit{PI} image is shown in Fig. \ref{fig:obs_PDI}. In our image, the binarity of GG\,Tau\,Ab1/2 is not resolved, therefore we refer to this component in the following as Ab. The image shows bright emission close to Aa and Ab, followed by a gap that is surrounded by the bright circumbinary ring.
\begin{figure}[t]
    \begin{centering}
    \includegraphics[width=0.47\textwidth]{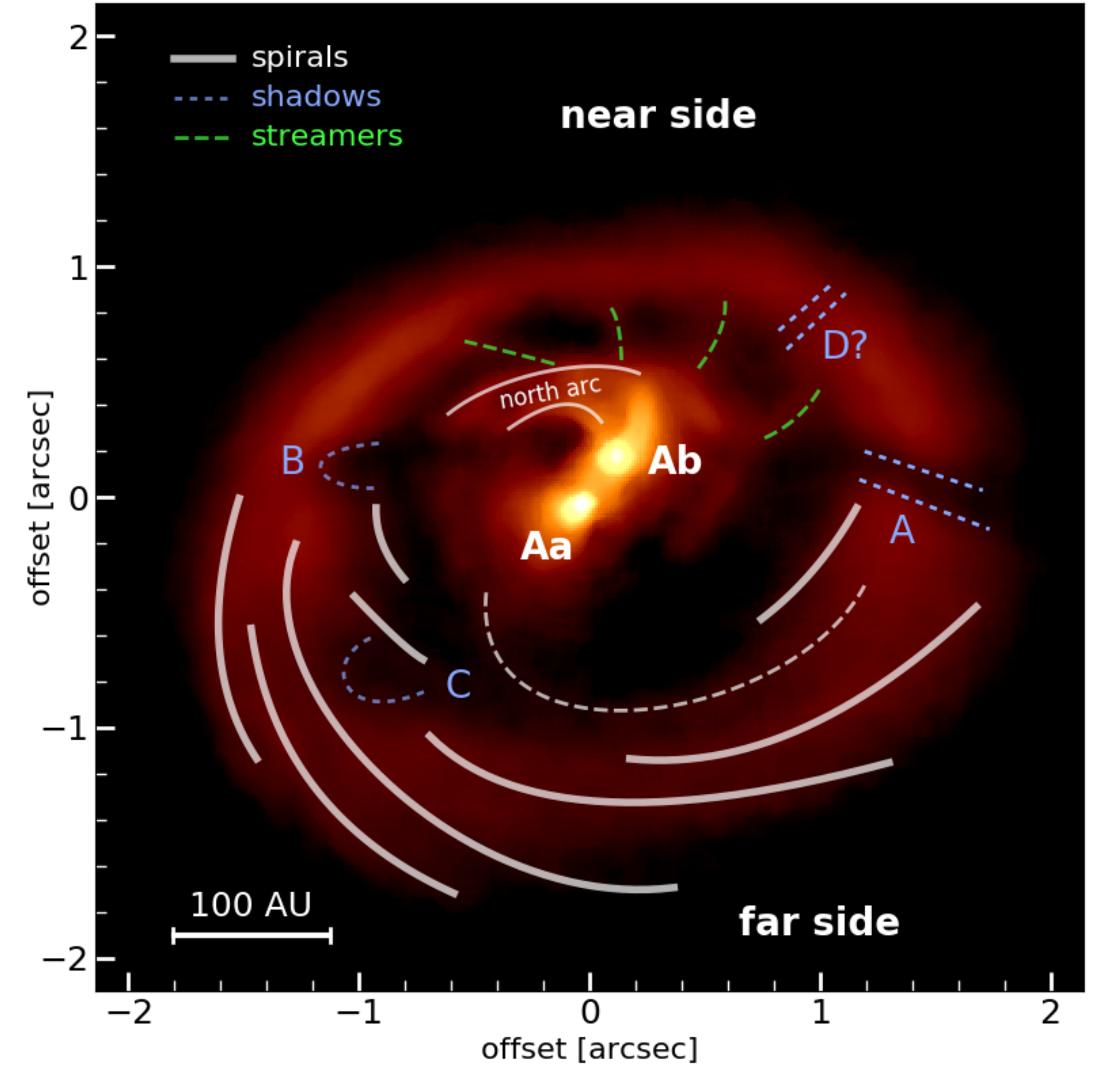}
    \end{centering}
    \caption{Schematic overview of the features in the outer circumbinary ring detected in our SPHERE \textit{PI} image. The image is centered on the location of GG\,Tau\,Aa and was smoothed for illustration purposes.
}\label{fig:outer_disk_features}
\end{figure}
The circumbinary ring is highly structured, with several shadowed regions, as well as several fine filament structures connecting the northern side of the ring with the close environment of the binary, and spiral structures in the southern disk region. Figure \ref{fig:outer_disk_features} presents a schematic overview of the detected features in the outer disk region. The following sections are dedicated to a detailed characterization of the different disk regions and categories of substructures.

%--------------------------------------------------------------------

\subsection{Inner region}\label{sec:inner_region}

\begin{figure}[bt]
    \begin{centering}
    \includegraphics[width=0.5\textwidth]{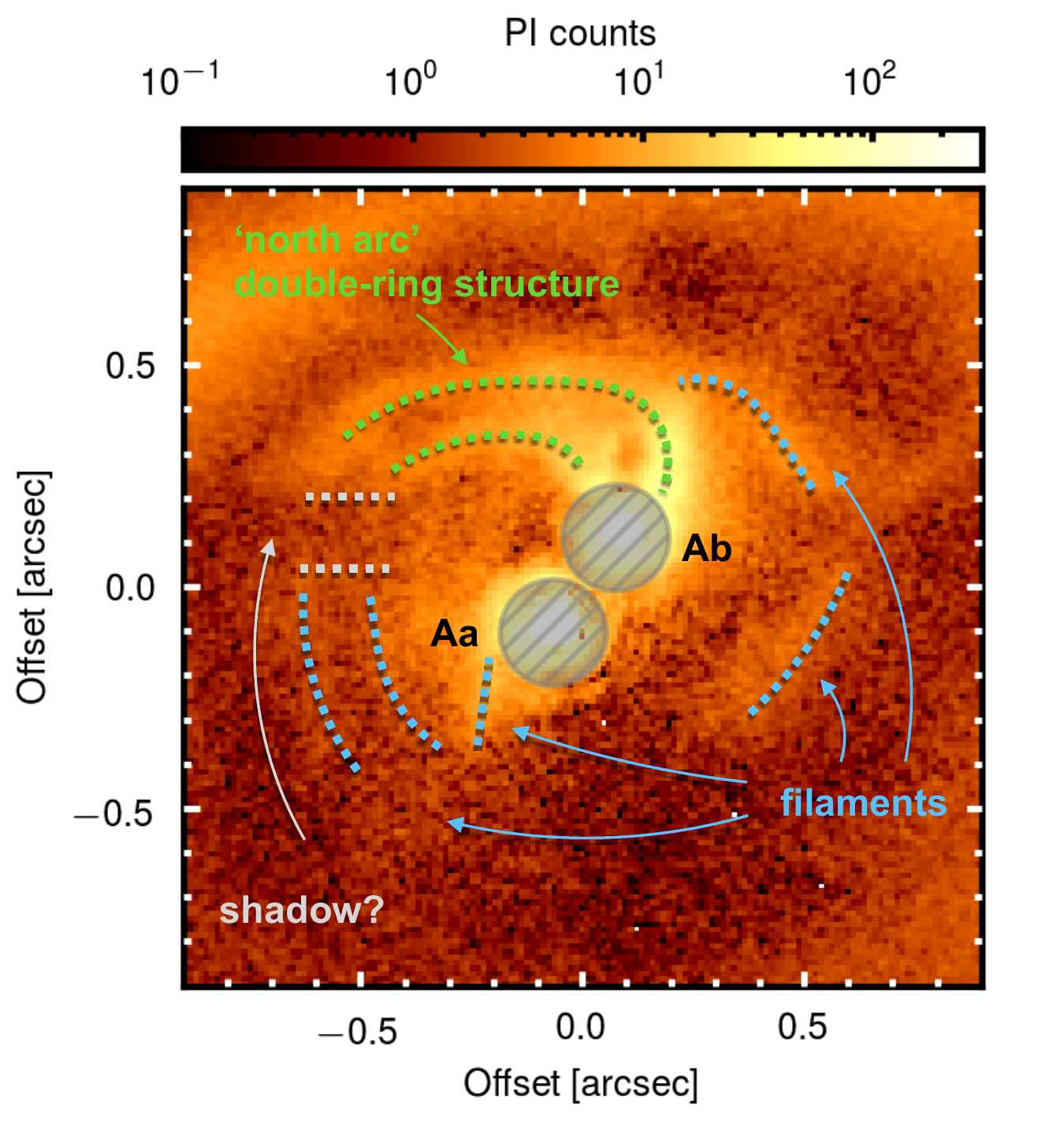}
    \end{centering}
   \caption{
   Zoom on the inner region after subtraction of the stellar polarization on Aa. The dotted lines highlight the detected features: the 'north arc', revealing a double-arc structure (green), several filaments (light blue) and a possible shadow lane (gray dashed). See Sect. \ref{sec:inner_region} for details. The immediate stellar environments ($<$\,120\,mas) are masked out. North is up and east is to the left.
}\label{fig:inner_region}
\end{figure}

Our final image, after correction for the instrumental polarization effects, reveals a residual unresolved polarized intensity signal at the locations of both Aa and Ab. We measure a linear polarization degree and angle of 0.33$\%$ and 37.1\degree \ at the location of Aa, and 1.12$\%$ and 8.7\degree \ at the location of Ab. A non-negligible amount of residual polarization can be interpreted as signal from unresolved circumstellar material such as a disk observed at nonzero inclination \citep[e.g.,][]{vanHolstein+20,Keppler+18,Garufi+20}. The circumstellar material around both components is confirmed by the measurement of non-negligible extinction \citep[$A_V$\,=\,0.3\,mag and 0.45 mag toward Aa and Ab, respectively; ][]{Hartigan+Kenyon03}, as well as accretion signatures from hydrogen-recombination lines and 10\,$\mu$m silicate features found at the location of both components \citep{White+99, Hartigan+Kenyon03,Skemer+11}. While we cannot make a statement about the inclinations of the disks from our measurements (except for excluding the case where the disks would be seen face-on and are circular symmetric: in this case, the polarized signal would cancel out), the measured angles of linear polarization indicate that the disks are oriented at position angles of $\sim$127\degree \ and $\sim$99\degree \ (i.e.,  perpendicular to the direction of linear polarization), respectively. We note that close to the stars, the radiation field is dominated by their individual illumination, and the contribution to the measured residual \textit{PI} from the respective other star can be neglected (see Sect. \ref{sec:app:scattering}). Observations at higher angular resolution and/or detailed modeling are required to better constrain the orientation of the circumstellar disks.

Any unresolved circumstellar material may create a halo of polarization signal around the star. Because this unresolved polarized signal can affect the analysis of the immediate circumstellar environments, we subtracted these polarized signals (i.e., the total intensity halo multiplied by the degree of polarization) individually for Aa and Ab.
   While subtracting the polarization signal of Aa slightly increases the contrast of the fine structures in the immediate stellar environment, subtracting the polarization signal of Ab instead blurs these structures. This can be explained by the fact that the measured polarization degree of Ab is somewhat higher than that of Aa. Therefore, subtracting the polarization signal of Ab adds an artificial polarization halo around Aa, which weakens the fine structures in its environment. Subtracting the (less strongly) polarized signal of Aa, however, does not noticeably affect the environment of Ab. Figure \ref{fig:inner_region} shows the resulting image after subtraction of the polarization signal of Aa with annotations of the detected features. The immediate stellar environments that are affected by the diffraction pattern are masked out. In all the images, the inner region appears highly structured, as highlighted in Fig.\,\ref{fig:inner_region} by the dotted lines. Most prominently, the `north arc', an extended structure to the northeast of Ab observed in previous scattered light images \citep[e.g.,][]{Krist+02,Krist+05,Itoh+14,Yang+17}, is clearly detected and appears in our SPHERE image to be composed of a double-arc structure at projected separations of $\sim$\,0.38\arcsec \ and $\sim$0.48\arcsec.
This double-arc structure may extend along the entire eastern side to the south, interrupted by a dark lane extending from Ab toward the east (see the dotted gray lines in Fig.\,\ref{fig:inner_region}). This dark lane seems to be connected to the shadow observed in the outer disk at a similar position angle (see Sect. \ref{sec:shadows}). We furthermore detect two additional filament structures northwest and southwest of Ab. It is unclear, however, whether they are related to the double-arc system on the eastern side. Finally, another filament is detected immediately southeast of Aa, pointing toward the south.

\begin{figure*}[htb]
    \begin{centering}
    \includegraphics[width=1.0\textwidth]{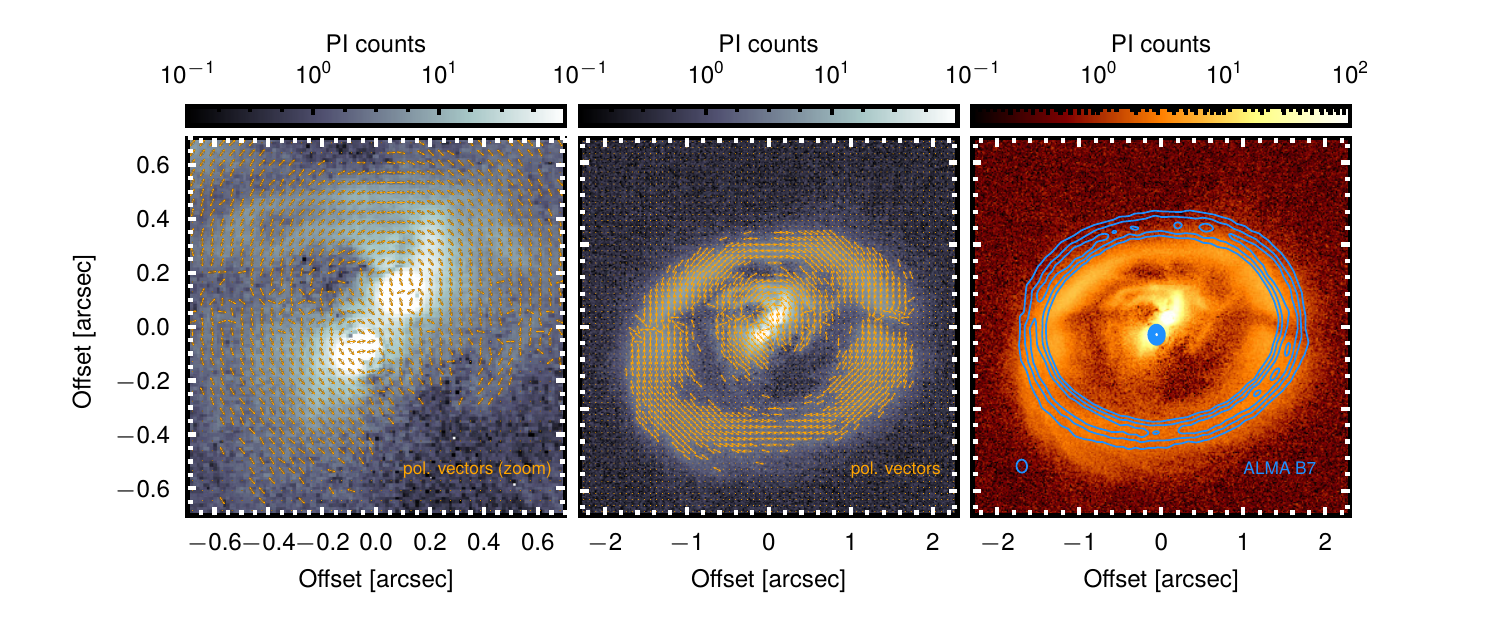}
    \end{centering}
   \caption{
   \textit{Left and center:} SPHERE \textit{PI} image with lines indicating the angle of linear polarization overplotted, showing two different fields of view (0.7\arcsec$\times$0.7\arcsec, 2.3\arcsec$\times$2.3\arcsec). The lines have an arbitrary length.
   For the computation of the polarization angles, we ignored bins at which the binned polarized intensity values $\leq$1.9. 
   \textit{Right:} polarized intensity image with ALMA Band 7 (0.9\,mm) continuum contours overplotted (blue). The ALMA observations were published in \cite{Phuong+20}. The ALMA image was registered such that the inner continuum emission, attributed to a circumstellar disk around Aa, coincides with the NIR position of Aa. Contours are shown at 20, 30,.., 80, 90\% of the peak intensity. The beam size is indicated in the lower left corner.
}\label{fig:pol_vectors}
\end{figure*}

Figure \ref{fig:pol_vectors} (left panel) shows the angles of linear polarization overplotted on the inner disk region. The polarization angles $\theta$ were calculated according to $\theta = 0.5 \times \mathrm{arctan}(U/Q)$, within bins of 3 pixels. 
Within the entire inner region, the polarization vectors appear to be generally aligned in azimuthal direction, as expected for light that is scattered off dust particles illuminated by a central source. 
Deviations from azimuthal polarization, as in the southwest from Aa, for example, may be due to the complex illumination pattern by the binary, or they might indicate multiple scattering events \citep[e.g.,][]{Canovas+15}. We note that the disk substructures we detected and highlight in Fig. \ref{fig:inner_region} cannot be explained by a potential interference of polarization vectors in the presence of two illumination sources, which might in principle lead to cancelling \textit{PI} out if the polarization vectors included an angle of 90\degree \ (see Sect.\,\ref{sec:app:scattering}).
This illustrates that small grains scatter light from the central illumination sources within a large region around the binary. \\

%--------------------------------------------------------------------

\subsection{Outer disk geometry}\label{sec:outer_disk}

\begin{figure}[bth]
    \begin{centering}
    \includegraphics[width=0.5\textwidth]{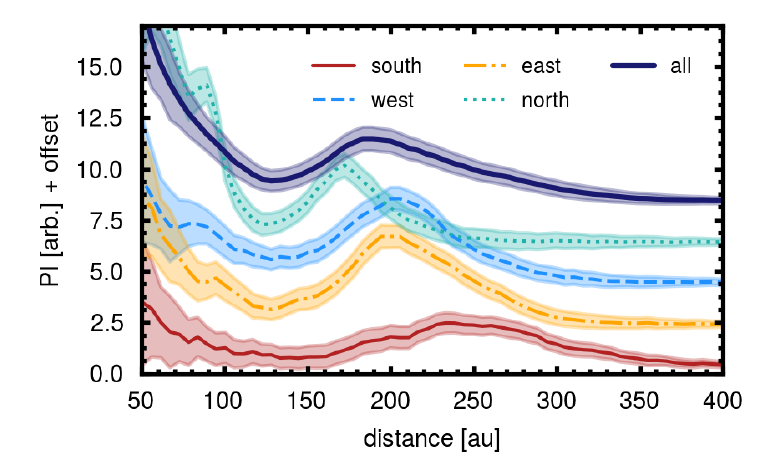}
    \end{centering}
   \caption{Radial disk profiles, taking into account the disk inclination of 37\degree. The profiles are drawn along the major (east, west) and minor (north, south) axes within an azimuthal cone of $\pm$20\degree \ around the corresponding axes, as well as averaged over all azimuths. The radial bin size is 3 pixels.} 
\label{fig:radial_profile}
\end{figure}
As in previous observations, the outer circumbinary disk appears as a large elliptical ring. The polarization angles in the center panel of Fig. \ref{fig:pol_vectors} (here calculated within bins of 6 pixels) show that also in the outer disk, the detected signal is overall well polarized in the azimuthal direction.
Only emission within two shadowed regions (shadows A and B, see Sect. \ref{sec:shadows}) appears to be less consistently aligned, owing to the lower signal-to-noise ratio. We note that while the circumbinary ring appears bright in (sub-)millimeter continuum observations \citep[e.g.,][]{Guilloteau+99,Dutrey+14,Tang+16,Phuong+20}, the region inside the ring reveals little to no signal at these wavelengths, except for an unresolved source at the location of Aa. This is illustrated by an overlay of the SPHERE image with the contours of the ALMA dust continuum at 0.9\,mm \citep{Phuong+20} in Fig.\,\ref{fig:pol_vectors} (right). This may imply that dust grains are mostly of small size inside the cavity, consistent with large grains being trapped in the outer circumbinary ring, while small grains, well coupled to the gas, can still enter the cavity \citep[e.g.,][]{Pinilla+12,DeJuanOvelar+13,Cazzoletti+17}. The comparison of the SPHERE and ALMA images also shows an obvious shift of the ring roughly along the disk minor axis, which is due to a projection effect related to the fact that the ALMA dust continuum traces the disk midplane, while the near-infrared SPHERE observations image the scattering surface of the disk. This is explained in detail in the following.

Figure \ref{fig:radial_profile} shows the radial deprojected profiles of the linear polarized intensity averaged along the major and minor axes, as well as averaged over the complete range of azimuthal angles. The polarized intensity along the major axis peaks around 200\,au.
Although the derivation of the profiles takes the projection by the disk inclination into account, the profiles along the near (north) and far (south) side of the minor axis appear very different: while the profile along the near side is quite peaked and peaks farther in than the major axis ($\sim$175\,au), the profile of the far side is much broader and peaks at a much larger distance ($\sim$250\,au). The different peak locations along the minor axis reflect a shifted geometric center of the ring because the ring is not geometrically flat, but has a non-negligible thickness. Similarly, the different profile shapes (broad versus peaked) are also connected to the geometrical thickness because the inclination of the disk allows us to see the inner rim of the southern (far) side, while for the north (front) side, the inner wall is hidden and only the upper surface is visible \citep[e.g.,][]{Silber+00,Krist+05}.

To quantify the outer ring geometry, we extracted radial profiles within azimuthal bins of 20\degree \ width. For each azimuthal bin, we determined the location of maximum brightness by fitting a polynomial function to the radial profile and then fitting an ellipse to the radial peak locations at all position angle bins. We find that the ring can be fit with an ellipse of eccentricity 0.64, a semimajor axis of 216\,au, and a position angle of 288\degree. The geometric center of the ellipse is offset by 32\,au toward the south from the assumed center of mass. These results compare well with the values found in previous scattered-light studies at similar wavelengths \citep[e.g.,][]{McCabe02}.
If the disk were geometrically flat and intrinsically circular, an eccentricity of 0.64 would imply an inclination of 39.7\degree. This value is slightly higher than the inclination of 37\degree$\pm$1\degree \ derived from (sub-)millimeter continuum observations \citep{Guilloteau+99, Andrews+14} because the geometric thickness of the disk affects the scattered-light observations \citep[e.g.,][]{Guilloteau+99,McCabe02,Krist+02}. The measured offset $\Delta$s of the geometric center of the ellipse from the assumed system barycenter can be used to constrain the scattering surface height $H_{\tau = 1}$ along the ellipse according to $H_{\tau = 1}(r)=\Delta s(r)/\mathrm{sin(}i\mathrm{)}$ \citep[e.g.,][]{deBoer+16}. Our measured offset $\Delta$s of 32\,au therefore corresponds to a scattering height of $\sim$53\,au at the inner edge of the ring ($\sim$200\,au). Because the scattering surface height typically traces layers at about 2-3 times the pressure scale height $H_p$, this would imply an aspect ratio of $H_p/R\sim$0.09-0.13, which compares well with constraints from other disks \citep[e.g.,][]{Villenave+19}.
We stress that this should only be considered as a rough estimate because azimuthal variations of the surface brightness, due to the azimuthal dependence of phase function and polarization degree, as well as the abundance of disk substructures such as shadows and spirals, may complicate a precise determination of the isophotes to which our ellipse was fit. 

Finally, a precise knowledge of the vertical thickness of the ring is required in order to determine the disk eccentricity from the scattered-light data.
However, optically thin millimeter observations indicate that the intrinsic eccentricity of the ring is rather low because the continuum, which traces the emission from the disk midplane and whose shape is therefore less biased by geometrical effects, can be well fit by an intrinsically circular model ring at the given angular resolution   \citep[beam major axes of 0.45\arcsec and 0.67\arcsec; ][]{Pietu+11,Andrews+14}. \\

\begin{figure*}[tb]
    \begin{centering}
    \includegraphics[width=0.9\textwidth]{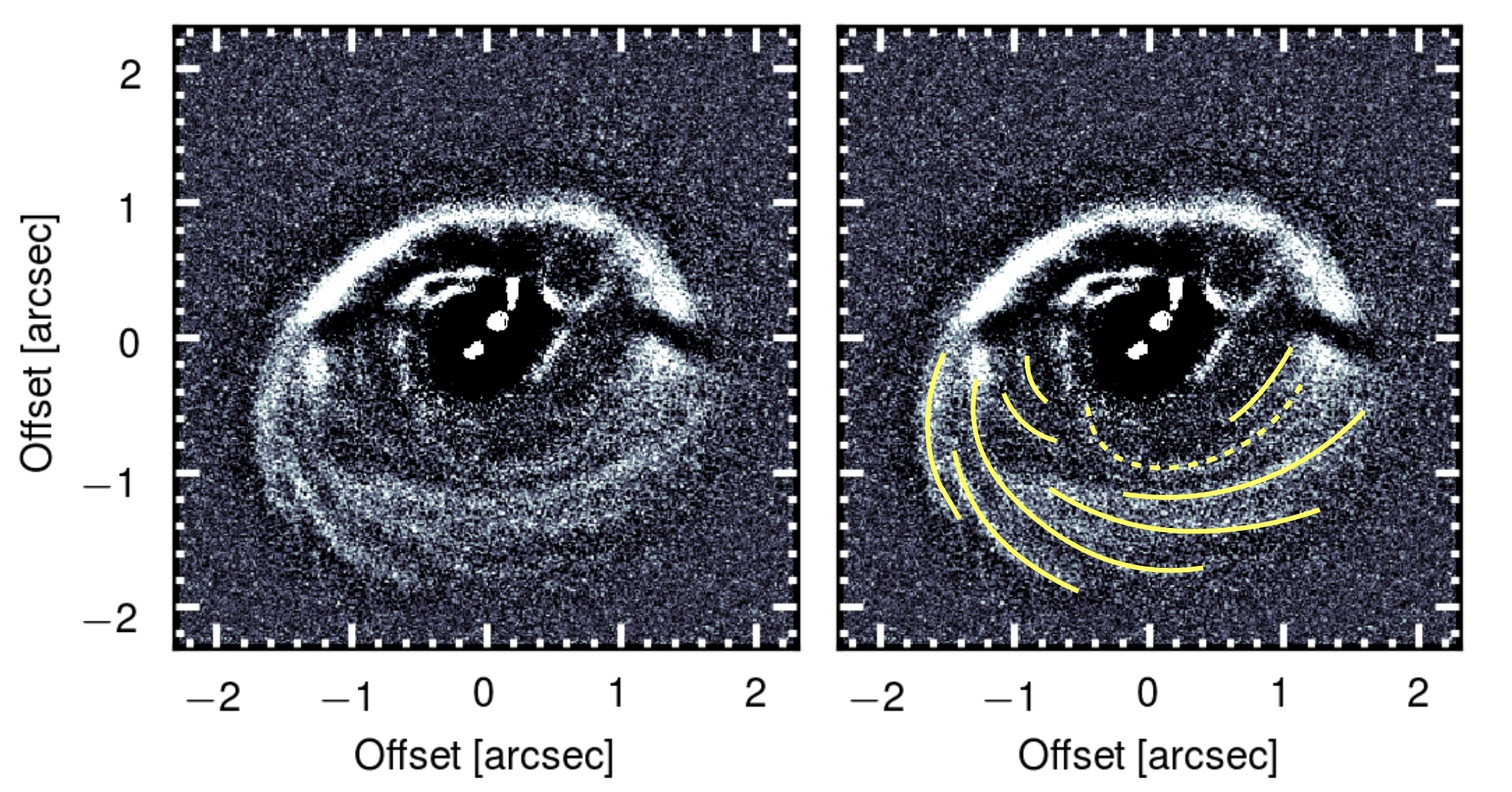}
    \end{centering}
   \caption{High-pass filtered PI image of GG\,Tau\,A (\textit{left}) with highlighted spiral structures (\textit{right}). The dashed line highlights the possible connection of the outer disk to Aa. 
}\label{fig:spirals}
\end{figure*}

%--------------------------------------------------------------------

\subsection{Streamers}\label{sec:streamers}
We detect four filament-like structures connecting the inner edge of the outer disk and the outer edge of the northern arc, as indicated in green in Fig. \ref{fig:outer_disk_features}. Some of these structures have previously been described as 'bridges' by \cite{Itoh+14} and \cite{Yang+17}.

In order to measure the position angles of these structures, we deprojected the image, assuming $i$\,=\,37\degree \ and $PA$\,=\,277\degree. The connecting points of the filaments at the inner edge of the outer disk are found at approximately $PA\sim$\,296\degree, 331\degree, 0\degree, and 36\degree \ (from west to east). The filaments are not aligned with the radius vector pointing toward the center of mass, but are tilted by increasing angles from west to east of $\sim$13\degree \ to 26\degree \ with respect to the radial direction.
The measured \textit{PA}s imply that the azimuthal spacing of the filaments is about 29\degree, 35\degree \ , and 36\degree. When we adopt an arbitrary uncertainty on the $PA$ measurement of 5\degree, this translates into a mean spacing of 33.3\,$\pm$\,2.9\degree. When we assume that the outer disk is in Keplerian rotation around a center of mass with 1.15\,\Msun, the azimuthal spacing of the filaments may imply that the filaments are launched by periodic perturbations occurring at the inner edge of the disk (180\,$\pm$\,20\,au) every 208\,$\pm$\,29 years.

The binary best-fit semimajor axis of 36.4\,au constrained by \cite{Koehler+11} (scaled here to 150\,pc) translates into an orbital period of about 205\,years, assuming a central binary mass of 1.15\,\Msun. The azimuthal spacing of the filaments would therefore be compatible with being triggered by a periodic perturbation occurring once every binary orbit, when the secondary passes at apocenter and comes closest to the disk edge. Interestingly, when we assume that the binary orbit is coplanar with the disk, the binary has just passed apastron \citep{McCabe02}. 

We interpret the filaments as accretion streams. Accretion streams close to the north arc have previously been suggested by continuum observations at 1.1\,mm \citep{Pietu+11}, as well as by the CO\,J\,=\,6-5 emission line, which show deviation from Keplerian rotation that may be compatible with infall motion \citep{Dutrey+14}. Furthermore, the $^{12}$CO gas distribution within the cavity shows a highly inhomogeneous structure consisting of several fragments \citep{Dutrey+14}. One of these CO fragments coincides with the location of the northern arc. As noted by \cite{Yang+17}, the entire northern arc may thus itself be part of a large accretion stream. \\

%--------------------------------------------------------------------

\subsection{Shadows}\label{sec:shadows}

\begin{figure}[tb]
    \begin{centering}
    \includegraphics[width=0.4\textwidth]{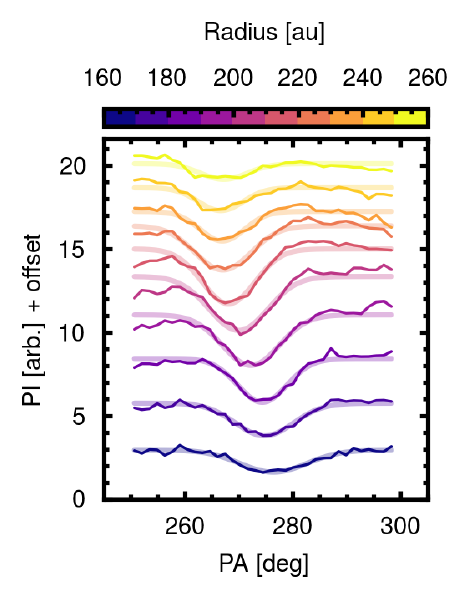}
    \end{centering}
    \caption{Azimuthal profiles of the western shadow at different (deprojected) radial bins between 160 and 260 au. The smooth, shadowed lines correspond to the best-fit Gaussian profiles, respectively.
}\label{fig:western_shadow_profiles}
\end{figure}
We detect three shadowed regions, known from previous scattered-light observations \citep[e.g.,][]{Silber+00,Itoh+14,Yang+17}, at \textit{PA} $\sim$275\degree, $\sim$86\degree, and $\sim$132\degree, and labeled A, B, and C in Fig. \ref{fig:outer_disk_features}. In addition, we detect a tentative fourth shadow, labeled 'D', at a \textit{PA} of about 320\degree \ and possibly related to a less prominent gap detected at a similar location (\textit{PA}$\sim$310\degree) by \cite{Krist+05}.

The most prominent shadow is a dark lane close to the western major axis of the disk (shadow `A'). To measure the shadow location, we deprojected the disk assuming an inclination of 37\degree, transformed the image into polar coordinates, and traced the azimuthal profile of the shadow in different radial bins. We then fit a Gaussian profile with negative amplitude to these profiles. At the inner and outer edge of the ring ($\sim$\,175\,au and $\sim$\,245\,au), we find the shadow center to be located at \textit{PA} of 274.8\degree \ and 266.7\degree, respectively. The tilt of the shadow is therefore about 8\degree. We furthermore measured the contrast of the surface brightness in polarized intensity within the shadow lane with respect to the disk just north of it, resulting in a contrast of about 2.6.

Using the Subaru datasets taken in 2001 January and 2011 September, \cite{Itoh+14} measured an anticlockwise rotation of the shadow of 5.9\degree \ and 4.9\degree \ between both epochs for the inner and outer disk edges, respectively. If the movement were linear in time, we would expect a further displacement by $\sim$2.5-3\degree \ between 2011 and our SPHERE dataset taken in 2016 November.
In order to verify the movement, we repeated our procedure of determining the shadow location on the total intensity frame of the 2011 Subaru dataset. We measure a shadow \textit{PA} of 274.4\degree \ and 268.1\degree \ at the inner and outer edge of the disk as defined above. Compared to the values we measured on our SPHERE dataset above (274.8\degree\ and 266.7\degree), we therefore cannot confirm a linear movement of the shadow between 2011 and 2016. The shadow positions instead appear to be stable.

It has been suspected that this western shadow may be cast by circumstellar material \citep[e.g., ][]{Itoh+14}, such as by an inclined disk around one of the binary components, as in the case of HD\,142527 \citep{Marino+15}. Shadow `B' (and the dark lane to the east of Ab detected in the inner region, see Sect. \ref{sec:inner_region}) may be just the east side of this same shadow \citep[see also ][]{Brauer+19}.
We can estimate the expected brightness contrast of the shadow lane with respect to the adjacent nonshadowed disk region under the hypothesis that one or two of the illumination sources are blocked by an optically thick inclined disk. 
A disk region that does not lie in any shadow is illuminated by all three stars, and it will therefore receive a total flux of $F_{\mathrm{tot}} = (1/4\pi) \times (L_{\mathrm{Aa}}/d_{\mathrm{Aa}}^2 + L_{\mathrm{Ab1}}/d_{\mathrm{Ab1}}^2 + L_{\mathrm{Ab2}}/d_{\mathrm{Ab2}}^2)$, where $d_{x}$ is the distance of component $x$ to the shadowed disk region. When one of the stellar components is surrounded by an inclined optically thick disk, this will cast a shadow on the outer disk, which will therefore only be illuminated by the two remaining sources.
Using the stellar luminosities as defined in Table \ref{table:stellar_properties}, and estimating a distance of Aa and Ab to shadow `A' of $\sim$181\,au and $\sim$156\,au, respectively (as measured on the deprojected image), we would expect a contrast of $\sim$2.4 for a disk around Aa, a contrast of $\sim$1.7 for a disk around Ab (i.e., a circumbinary disk around Ab1/2), and a contrast of $\sim$1.4 and $\sim$1.2 for a disk around Ab1 and Ab2, respectively.  The measured contrast of 2.6 from our SPHERE data would therefore favor the shadow to be cast by an inclined disk around Aa or Ab, rather than around Ab1 or Ab2.

\cite{Min+17} have developed an analytical description with which the orientation of an inner shadow-casting disk can be derived from measuring the orientation of the shadows cast on the outer disk. We repeated the same procedure for GG\,Tau\,A, assuming that the shadow is cast by a disk around either Aa or Ab. For this purpose, we measured the position angle of the line connecting the two shadows of about 90\degree, and the vertical (projected) offset of this line of 21.2\,au and -9.7\,au from Aa and Ab, respectively. Inserting these values into equations 7 and 10 of \cite{Min+17}, we obtain a disk position angle of about 90\degree \ for the shadow-casting disk for both cases. Assuming an outer disk aspect ratio of 0.1-0.15, and assuming that the scattering surface is found at about 2-3 times the pressure scale height, we furthermore find an inclination of $\sim$72\degree-81\degree \ if the disk were found around Aa, and an inclination of $\sim$96\degree-100\degree\ if it were found around Ab. Considering the outer disk inclination of 37\degree, the misalignment of a disk around Aa and Ab would then be $\sim$35\degree-44\degree \ and $\sim$59\degree-63\degree, respectively.

Recently, \cite{Brauer+19} have investigated the effect of circumstellar disks around the binary components on the brightness distribution within the circumbinary ring using radiative transfer modeling. In one of their setups, they simulated an inclined circumstellar disk around Ab2 (while keeping a coplanar disk around Aa). In this case, their simulations were able to reproduce a sharp shadowed lane at the location of shadow `A', as well as a symmetric eastern shadow (corresponding to shadow `B'), although they found it to be shallower in brightness contrast than in the observations. We suggest here that an inclined disk around Ab (i.e., a circumbinary disk around both Ab1 and Ab2) or around Aa would be more compatible with the measured contrast.

%--------------------------------------------------------------------

\subsection{Spirals}

We detect multiple spiral structures in the southern part of the disk. For an improved identification, we processed the image by a high-pass filter, that is, we convolved the image with a Gaussian filter ($\sigma$\,=\,9\,px) and subtracted it from the original image. The spiral structures are clearly seen in this image (Fig. \ref{fig:spirals}). Interestingly, one spiral arm is tentatively found to cross the gap, and if confirmed, connects the southwestern circumbinary ring to the immediate circumstellar environment of Aa (see the dashed yellow line in Fig \ref{fig:spirals}, right).
Thin filaments in the southeast disk have previously been suggested from the observations by \cite{Krist+05}, who interpreted these structures as possible signs of binary-disk interactions. Furthermore, \cite{Tang+16} and \cite{Phuong+20_spirals} found at an angular resolution of $\sim$0.3-0.4\,\arcsec, that the radial distribution of CO brightness in the outer disk exhibits several spiral structures. \\

%-----------------------------------------------------------------------
%-----------------------------------------------------------------------

\section{Modeling}\label{sec:modeling}
We performed hydrodynamical simulations in order to model the system and its evolution. The main goal was to verify whether the binary might be qualitatively responsible for the observed gap size and features within the circumbinary ring.

%-----------------------------------------------------------------------

\subsection{Hydrodynamical model setup}\label{sec:hydro}
We carried out hydrodynamical simulations of the gas disk using the GPU version of {\tt\string PLUTO} \citep{Mignone2007} by \cite{Thun2018}. The simulations were 2D and isothermal. We used a polar radially logarithmic grid ranging from one binary semimajor axis ($a_{\mathrm{bin}}$, 35\,au) to 40\,$a_{\mathrm{bin}}$ (1400\,au) with 684 cells in radial and 584 cells in azimuthal direction.
Because the separation of Ab1 and Ab2 \citep[$\sim$5 au; ][]{diFolco+14} is smaller than the inner edge of the circumbinary ring ($\sim$200 au), we considered Ab1 and Ab2 together as a single component, Ab, and the entire system was treated as a binary.
The binary components Aa and Ab were assumed to have masses of 0.75\,\Msun\ and 0.67\,\Msun, implying a mass ratio of 0.89, similar to the mass ratio of 0.77 derived in Sect. \ref{sect:stellar_parameters}. As shown in \cite{Thun2018}, minor changes in the mass ratio of the binary affect the disk dynamics only very slightly.
The binary orbit was set to have a semimajor axis of 35\,au and an initial eccentricity of 0.28, consistent with the observations \citep{Koehler+11}. Furthermore, the binary orbit was assumed to be coplanar with the circumbinary disk plane. 
We ran two different models that differed only in the adopted radial temperature profile. In the first model, we considered a temperature profile constrained by the $^{13}$CO molecule \citep{Guilloteau+99}, tracing the disk surface temperature, and in the second model, we applied a temperature profile constrained by the dust continuum \citep{Dutrey+14}, tracing the midplane temperature,
\begin{align}
    T_{\mathrm{surface}} &= 20\,K \cdot \frac{300\,\mathrm{au}}{R} \\
    T_{\mathrm{midplane}} &=  13.8\,K \cdot \frac{200\,\mathrm{au}}{R.}
\end{align}
By considering these two different temperature profiles, which are sensitive to the warm disk surface and to the cool midplane, respectively, we covered the two limiting cases.
The aspect ratio $h = H/R$ of the disk was determined by the sound speed $c_s$ and Keplerian orbital frequency $\Omega_k$, and therefore results from the assumed temperature profile as follows: \\
\begin{align}
    h &= \frac{c_s}{{\Omega_k}R} = \sqrt{\frac{k_B}{G M_{\mathrm{bin}} \mu m_\mathrm{p}}} \cdot \sqrt{T R},
\end{align} 
 with $M_{\mathrm{bin}}$ the binary mass, $\mu$=2.3 the mean molecular weight, $m_P$ the proton mass, and $R$ the radial distance from the system barycenter in the disk plane. With our chosen temperature profile, we obtain a constant aspect ratio corresponding to the following values:
\begin{align}
    h_{\mathrm{surface}} &\approx 0.15 \\
    h_{\mathrm{midplane}} &\approx 0.11.
\end{align}

The initial surface density follows a power law $\propto$\,R$^{-1.5}$ normalized in such a way that the total disk mass amounts to 10$\%$ of the binary mass (0.14\,\Msun). As the inner $3\, a_{\mathrm{bin}}$ of the disk are unstable, the initial density profile inside of $2.5\, a_{\mathrm{bin}}$ exponentially decays to $e^{-1}$ of the smooth profile within $0.1\, a_{\mathrm{bin}}$. The boundary conditions of the simulations were defined as in \cite{Thun2018}.
We simulated the gas content of the disk assuming an $\alpha$ viscosity with a constant Shakura-Sunyaev parameter of $10^{-3}$ throughout the disk.

The computational time needed to reach the actual disk structure from the initial power-law profile can be long \citep{new_paper}. To ensure a feasible time step for the grid code, we did not include the stars themselves in the simulation domain, but the inner grid boundary was set to a radius of $1\, a_{\mathrm{bin}}$ (35\,au) and we added the binary as n-bodies inside the domain to create the potential, using a gravitational softening parameter of 0.6 \citep[see ][]{new_paper}.
As discussed in \cite{new_paper}, such an inner boundary does not change the dynamics of the circumbinary disk or gap width. The outer disk edge is an open boundary that assumes a continuation of the power-law disk. We note that the simulations do not take GG\,Tau\,B into account, which is observed at a projected separation of about 1400\,au from GG\,Tau\,A. Because this outer companion may accrete from and/or truncate the outer parts of the disk \citep[see, e.g.,][]{Beust+Dutrey06}, it is therefore possible that the density in the outer parts of the disk is overestimated in the simulation. We ran both models for 28\,000 binary orbits ($\approx$\,4.9\,Myr).

%-----------------------------------------------------------------------

\subsection{Postprocessing of hydrodynamical simulations}

To investigate the appearance of our simulated disks in scattered light, we generated images in polarized intensity using the  radiative transfer code RADMC-3D \citep{Dullemond+12}. We included a radiation field from two stellar components with luminosities of 0.44\,\Lsun \ and 0.20\,\Lsun and temperatures of 3900\,K and 3400\,K, respectively.
In order to generate a 3D view from the simulated disk, we expanded the 2D surface density distribution resulting from the hydrodynamical simulations along the vertical axis, assuming a Gaussian density distribution with constant aspect ratios of 0.15 and 0.11, consistent with the assumed temperature laws in the simulations (see Sect. \ref{sec:hydro}). We assumed the dust to be well mixed with the gas. This is a valid assumption because at 1.67\,\micron, the scattered light is dominated by micron-sized dust grains, which are well coupled to the gas.  We thus assumed the dust density distribution to be identical to that of the gas, scaled by a factor of 0.01, which corresponds to a typically assumed dust-to-gas ratio of 1 to 100 in protoplanetary disks.

We assumed the dust number density $n$ as a function of grain size $a$ to follow a power law of the form $n(a) \propto a^{-3.5}$. The grains were considered to be distributed between sizes of 0.005\,\micron \ and 0.5\,\micron, as assumed in the modeling efforts by \cite{Brauer+19}. We assumed that 5\% of the total dust mass is contained within this population of small grains, corresponding to a fraction of $5\times10^{-4}$ of the total disk gas mass.
Our dust mixture was composed of 70\% astronomical silicates \citep{Draine+03} and 30\% amorphous carbon grains \citep{Zubko+96}. 
We computed the Stokes Q and U frames at 1.67\,\micron, taking the observed inclination and position angle of the disk into account. The simulations were run using 10$^{8}$ photon packages in order to obtain images with high signal-to-noise ratios. Finally, we convolved our images with a Gaussian kernel with an FWHM of 43\,mas.

%-----------------------------------------------------------------------
%-----------------------------------------------------------------------

\subsection{Modeling results and comparison to observations}

\begin{figure*}
    \includegraphics[width=1.05\textwidth]{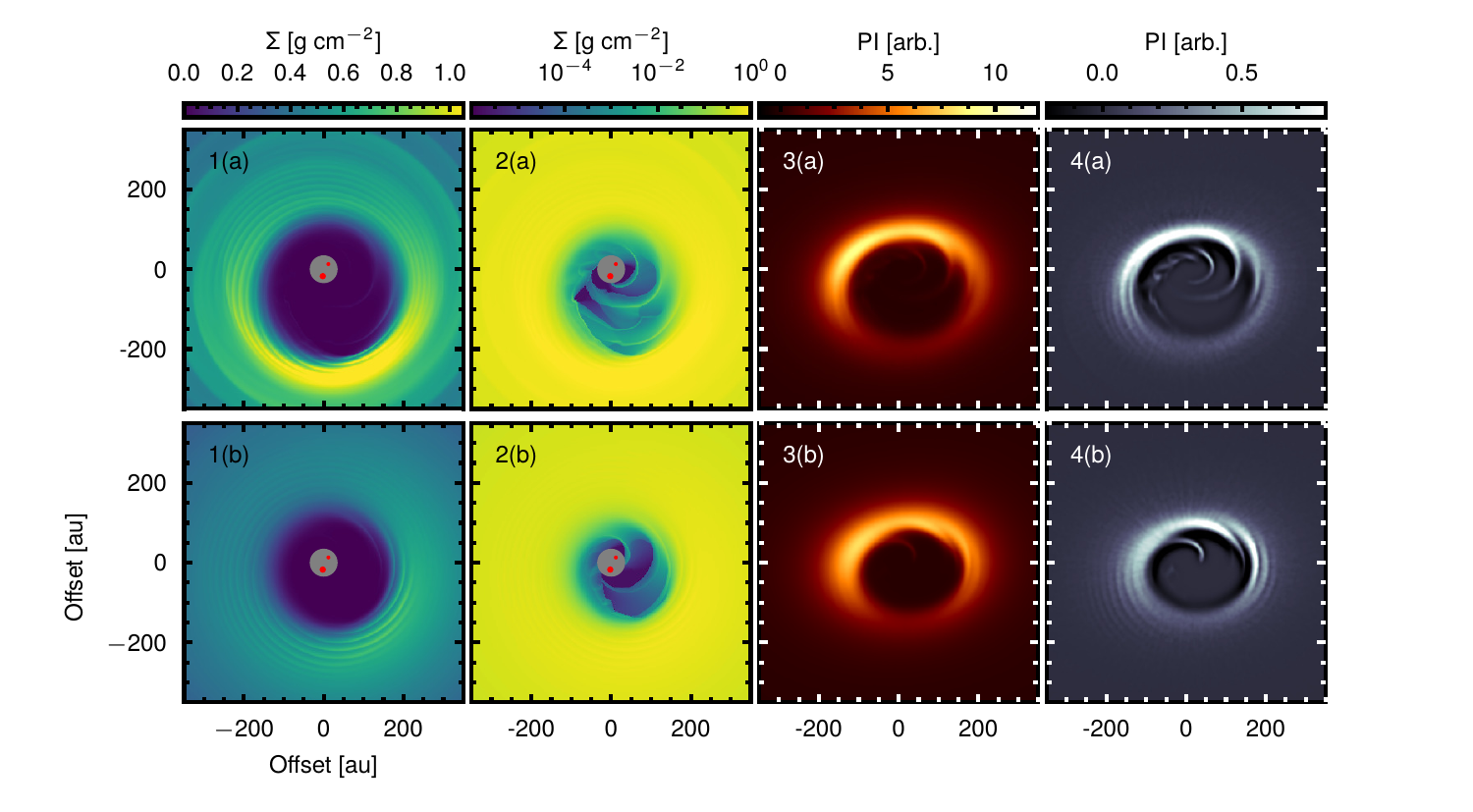}
    \caption{Surface density output of our simulations in linear (left column; 1 (a) and (b)) and logarithmic (center left column; 2 (a) and (b)) color stretch. The center right column (3 (a) and (b)) shows the simulated polarized intensity images evaluated at 1.67\,\micron. This image was calculated after inclining and orienting the disk as in the observations. The right column (4 (a) and (b)) shows the polarized intensity image of the center right column, processed with a high-pass filter.
    In each column, the panel in the first row (a) corresponds to the model with h=0.11, the panel in the second row (b) to h=0.15.}
\label{fig:sim_both}
\end{figure*}
\begin{figure}
    \includegraphics[width=0.5\textwidth]{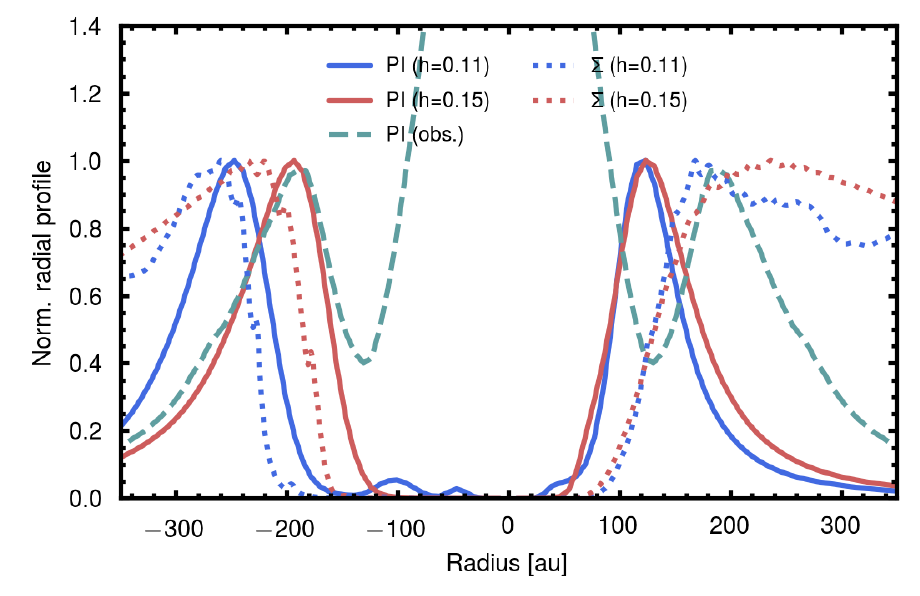}
    \caption{Radial profiles along the disk major axes (within a cone of $\pm$30\degree) of the two models drawn from the gas surface densities (dotted) and the deprojected, simulated polarized intenisty images (solid lines). As a comparison, the radial profile of the observed, deprojected disk averaged over all azimuths is plotted (dashed). } 
\label{fig:radial_profiles_sim}
\end{figure}

Both models ran for $28\,000$ binary orbits ($\approx$\,4.9\,Myr), during which the system reached a converging configuration of a stable sized, eccentric, precessing cavity around the binary and a stable circumbinary disk after about $10\,000$ binary orbits ($\approx$\,1.7\,Myr).
In the simulations, mass is constantly accreted onto the stars through accretion streams. Therefore the disk looses a fraction of about $5.2\cdot10^{-6}$ of its mass per binary orbit (or $3.0\cdot10^{-8}$ per year). As a result, the initial disk mass of 0.14\,\Msun \ has decreased at the end of the simulation to $86\%$ of its initial value (0.12\,\Msun). This is in excellent agreement with the disk mass constraints from observations \citep[$\sim$0.12 \Msun, ][]{Guilloteau+99}. We did not take the accretion onto the stars into account as it is not resolved in the domain.

The final gas density distributions for both simulations are displayed in Fig. \ref{fig:sim_both} (left and center left columns). They show evidence of large cleared inner regions. The gap in the gas is found to have a semimajor axis (defined as the location where the gas density has decreased to 10\,\% of its peak value) of $4.77\,\mathrm{a_{\mathrm{bin}}}$ (167\,au) and an eccentricity of 0.34 in the case of the midplane temperature ($h=0.11$), and a semimajor axis of $3.85\,\mathrm{a_{\mathrm{bin}}}$ (135\,au) and an eccentricity of 0.25 in the case of the (higher) surface temperature ($h=0.15$).

The surface density shows an azimuthal asymmetry, with the density peaking in direction of the disk apocenter. The reason is that the gas velocity is slowest at these locations, leading to an enhancement of material in these regions.
Fig.\,\ref{fig:sim_both} (left column) shows that the circumbinary ring is structured by numerous tightly wound fine spirals. 
Furthermore, the logarithmic color stretch for the surface density (Fig. \ref{fig:sim_both}, center left column) reveals the structure of material flow through the cavity. Spiral streams occur in the simulation, periodically driven by the circumbinary rotation, accelerating the close-by infalling material.
Our simulations show regularly stripped-off material streams from the outer disk, similar to the observations, while the exact morphology and orientation of the filaments is not reproduced.
These differences may be related to the fact that we do not know the exact initial conditions of the system, with some parameters such as its mass related to some uncertainty.
Another possible caveat in the simulations is the fact that we did not simulate the direct circumstellar material, but the simulation domain was cut inside of about 35\,au. The presence of material in that region (such as the `northern arc') may affect the flow dynamics and dust morphology within the cavity. Furthermore, it may affect the morphology of the material flow that Ab itself is a binary.

The postprocessed polarized intensity images are shown in the center right column of Fig. \ref{fig:sim_both}. The intensity also shows clear azimuthal variations here. Because the disk is optically thick in the near-infrared regime, the azimuthal dependence of the large-scale surface brightness is not sensitive to the surface density, but to the dust phase function and polarization degree.  As expected, the near side is significantly brighter than the far side.
The simulated polarized intensity images also show substructures within the circumbinary ring. While the contrast of the spirals in the circumbinary ring appears faint, they become very well visible when the images are treated with a high-pass filter, similarly to the observations (Fig. \ref{fig:sim_both}, right column). We note, however, that the simulated view of the disk in scattered light may be biased by our simplified treatment of the vertical structure of the disk.

Figure \ref{fig:radial_profiles_sim} shows the radial profiles of the simulated gas surface densities (dotted blue and red lines) along the disk major axes. We find a disk semimajor axis (defined as the distance  where the profile peaks) of about 215-230\,au. Assuming that the large dust particles traced by millimeter observations are being trapped at the location of maximum gas density, these findings are well comparable with the observations: using the optically thin continuum emission between 1.3\,mm to 7.3\,mm,  \cite{Andrews+14} observed the continuum to peak at about 250\,au.
Figure \ref{fig:radial_profiles_sim} also shows the radial profiles of the simulated deprojected polarized intensity images along the disk major axes (solid blue and red lines). In each of the cases, the polarized intensity profile peaks slightly ahead of the gas density. This can be explained by the fact that the peak of the scattered light profile traces the location of the inner wall of the ring, where illumination is strongest, and not directly the dust density distribution. The semimajor axes of the disk in the polarized intensity images are measured to be 180\,au and 160\,au, respectively. This is slightly shorter than the location of the peak of the mean (i.e., averaged over all azimuths) deprojected radial profile of the observed \textit{PI} image ($\sim$190\,au). One reason might be that the slope of the inner edge of the gas disk may in reality be somewhat sharper than in the simulations, which might be connected to the exact value of the binary eccentricity \citep[e.g.,][]{Miranda+17}, or to other disk properties such as the assumed temperature profile, density, distribution, and viscosity.
Furthermore, the rim location inferred from the scattered light observations may be overestimated because of possible shadowing from one (or several) circumstellar disks around the three individual components \citep{Brauer+19}.

Finally, the simulated gap cleared by the binary becomes eccentric, with mean eccentricity values of $\sim$0.2-0.3. As noted in Sect. \ref{sec:outer_disk}, it is difficult to extract reliable information about the disk eccentricity from the scattered-light observations, but the (sub-)millimeter observations indicate that the eccentricity of the disk is probably rather low \citep{Guilloteau+99,Andrews+14}. This might indicate a lower disk viscosity than assumed in our simulations, as discussed in Sect. \ref{sec:discussion_gap}. 

\begin{figure}[tb]
    \begin{centering}
    \includegraphics[width=0.50\textwidth]{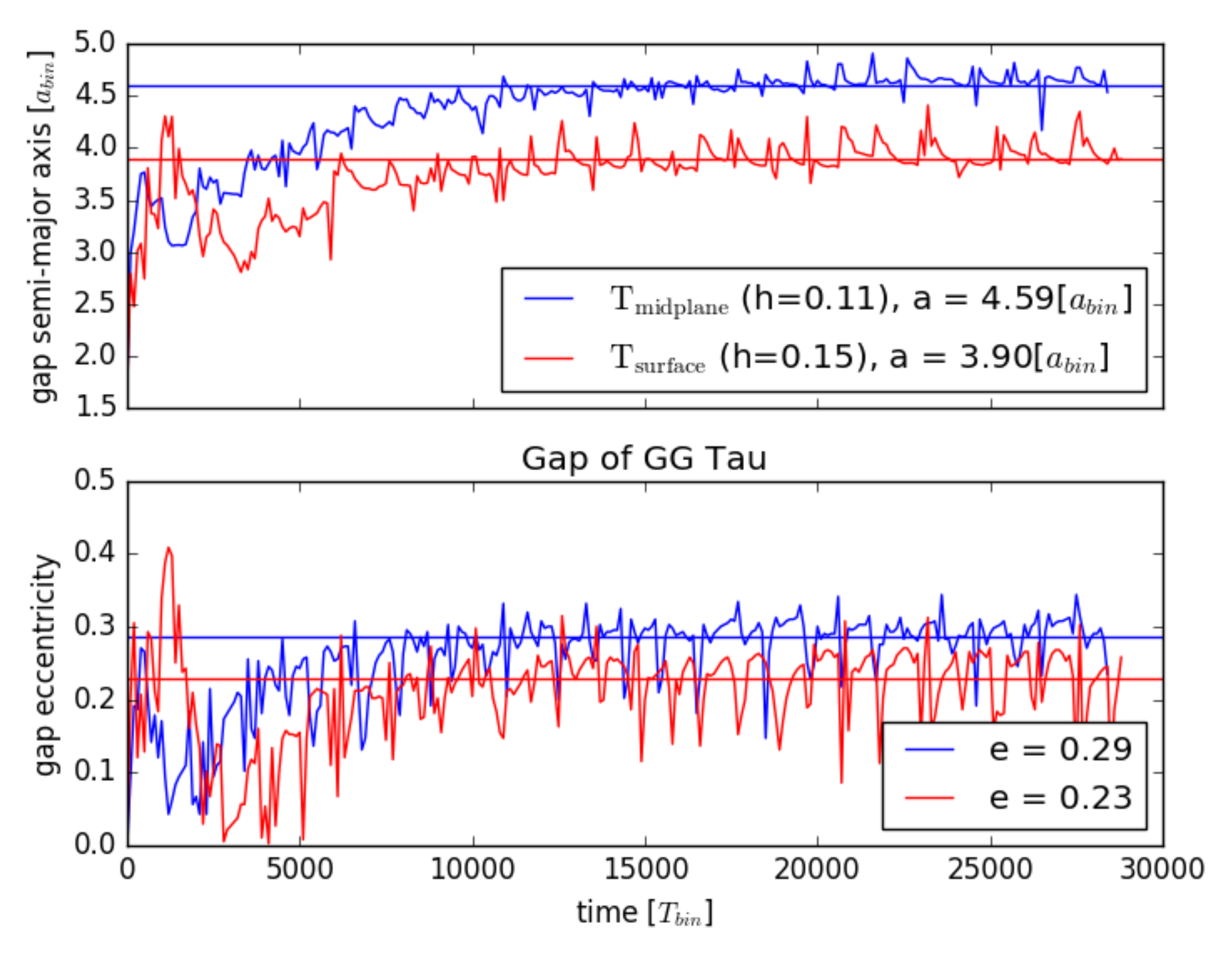}
    \end{centering}
   \caption{Size and eccentricity evolution of the cavity around GG\,Tau\,A created by the PLUTO simulations for the midplane temperature and surface temperature of the disk. $a_\mathrm{bin}$ is 35\,au and the initial disk mass is 0.1 binary masses.} 
\label{fig:sim_gap}
\end{figure}

%-----------------------------------------------------------------------
%-----------------------------------------------------------------------

\section{Discussion}\label{sec:discussion}

\subsection{Accretion streams within the circumbinary gap}
We interpret the filaments detected in our observations that we described in Sect. \ref{sec:streamers} as accretion streams. According to theoretical models, circumbinary accretion is thought to proceed onto the stars from the outer circumbinary disk through accretion streams, which are repeatedly torn off at the inner edge of the disk near the apocenter of the binary orbit. This is consistent with what is seen in our simulations. Such a phase-dependent pulsed accretion process has been seen in numerous theoretical studies \citep[e.g., ][]{Guenther+Kley02, Dunhill+15, Duffell+19}. Time-variable spectroscopic signatures of accretion activity (e.g., through hydrogen recombination lines) correlated with the binary orbital phase have been observed in some tight (spectroscopic) binary systems \citep[e.g.,][]{Mathieu+97,Kospal+18}.
While the detection of pulsed accretion is usually restricted to very tight systems (because of the restricted time base), the detection of periodic streamers in GG\,Tau\,A, if confirmed, would be the first indication of such processes in a wider system. 

The hypothesis that the filaments in GG\,Tau\,A indeed trace accretion streams fits the overall picture well. 
With large grains being trapped in the outer circumbinary disk, the detected streamers refill the immediate environment of Ab predominantly with gas and small grains, which is compatible with the strong silicate feature observed at the location of GG\,Tau\,Ab \citep{Skemer+11}. The formation of large, massive circumstellar disk(s) around Ab1/2, however, may be inhibited by its binary nature despite continuous replenishment of material, which could explain the nondetection of millimeter flux at the location of Ab \citep{Dutrey+14}.

%-----------------------------------------------------------------------

\subsection{Spiral structures as imprints of binary-disk interaction}

Our SPHERE observations show several spiral structures in the southern disk region. Our simulations show that this is an expected outcome of binary-disk interactions and is also consistent with other modeling efforts, which do show that the generation of spiral density waves is a common result of binary-disk interaction, in particular, for cases where the binary orbit has nonzero eccentricity \citep[e.g.,][]{Miranda+17,Price+18}. Observations of circumbinary disks have brought observational evidence of such spiral structures in these systems \citep[e.g., ][]{Avenhaus+17, Monnier+19}. In addition, large accretion streams, such as the tentative connection from the southern disk to the primary, Aa, are also expected from simulations \citep[e.g.,][]{Moesta+19}. In this respect, the detected spiral features agree well with our expectations from simulations of circumbinary disks, where the binary orbit has moderate eccentricity. We note that in addition to this, the \textit{\textup{external}} binary-disk interactions with GG\,Tau\,B (projected separation of $\sim$\,1500\,au) might also be able to trigger spiral waves in the GG\,Tau\,A disk \citep[e.g.,][]{Dong+16}. This scenario may be addressed by future work.

However, theoretical models have shown that in addition to binary-disk interactions, several different processes can also drive the generation of spirals in disks, such as a low-mass companion on an orbit inside or outside of the disk \citep[e.g.,][]{Dong+15}, gravitational instability \citep[e.g.,][]{Dong+15_gravinstab,Dipierro+15,Nelson+16,Meru+17}, or a combination of both \citep[e.g.,][]{Pohl+15}, as well as temperature fluctuations as a result of shadowing by a warped or misaligned inner disk \citep[e.g.,][]{Montesinos+16}.
In order to discern between the companion and gravitational instability scenario, observations at comparably high resolution of the dust continuum, probing the midplane of the disk, are required \citep[e.g.,][]{Rosotti+20}. However, we regard the last scenario as rather unlikely because in this case, the spiral arms would be expected to diverge from a location close to where the scattered-light shadows are located. In contrast, several spiral arms seem to rather originate from a point located on the outer ring  at a \textit{PA} of $\sim$120\degree. Interestingly, at this \textit{PA} (but slightly outward of the near-infrared peak emission, at radial distances of $\sim$215-270\,au), an asymmetric structure within the CO distribution has been found, showing evidence of a significantly increased temperature \citep{Dutrey+14,Tang+16}. This so-called `hot spot' was interpreted as an area with locally enhanced density and temperature, heated by a possible embedded planet at the formation stage \citep{Phuong+20_spirals}. While we still consider binary-disk interaction as the most obvious driving force for the spirals observed in the scattered light, a possible connection or interference with this hypothetical forming body needs to be investigated with complementary observations.

%-----------------------------------------------------------------------

\subsection{Gap size as a result of binary-disk interactions}\label{sec:discussion_gap}

Our simulations show that a binary with a semimajor axis of 35\,au is able to create a gap in a coplanar disk with a size that is comparable to our observations.
Our inferred gap sizes (4.8\,a$_{\mathrm{bin}}$ and 3.9\,a$_{\mathrm{bin}}$ for the two temperature regimes) agree with previous studies of other systems, in which cavity sizes ranging from three to six binary separations were found \citep[e.g.,][]{Thun2018}.

We note that our gap estimates are significantly larger than those derived by previous studies of the GG Tau A circumbinary ring. 
As an example, considering the best-fit astrometric solution of the binary under the assumption that the orbit is coplanar with the outer disk (a$_{bin}\sim$36\,au), \cite{Beust+Dutrey05} predicted a gap size of 2-3.3 a$_{bin}$, which they noted to be obviously incompatible with the observations.
 This apparent discrepancy between observed and simulated gap sizes was confirmed by the hydrodynamical simulations of \cite{Cazzoletti+17}, who tested the binary-disk coplanar case considering different disk temperature profiles and a range of values for the viscosity. The authors found the simulated gas distribution to peak at radial distances smaller than $\sim$160\,au, which contrasts with the observed millimeter continuum peak at about 250\,au. As a possible solution of this discrepancy, \citet{Beust+Dutrey05} proposed to drop the assumption that the binary orbit was coplanar with the disk. In this case, the most plausible orbit has a semimajor axis of $\sim$65\,au, an eccentricity of 0.44, and a disk-orbit misalignment of about 25\degree \ \citep{Koehler+11}. 
This latter scenario, a binary on a wide, disk-misaligned orbit was tested with hydrodynamical simulations by \cite{Aly+18}. The authors found that they were indeed able to reproduce the observed gap size, assuming a binary separation of $\sim$60\,au and a binary-disk misalignment of $\sim$30\degree. 

The differences between these earlier estimates and our own gap values are probably mainly due to a difference in timescales.
While our simulations were run for 28\,000 orbits, previous studies such as those of \cite{Nelson+16}, \cite{Cazzoletti+17} and \cite{Aly+18} stopped their simulations after about 1000-2000 orbits or fewer, and they therefore studied earlier stages of the disk evolution to define the gap size and eccentricity. Because the disk evolution starts from an azimuthally symmetric density distribution, our simulations show that the disk crosses meta-stable symmetric states between 1100 and 1700 orbits. This is illustrated in Fig. \ref{fig:cazo}, which shows the evolution of the gap semimajor axis and gap eccentricity over the first 20\,000 orbits. However, this meta-stable state is an artifact of the setup and does not correspond to the convergent behavior of a circumbinary disk because the binary will excite the disk to eccentric motion. The disk will eventually evolve to a larger, more eccentric, stable gap, as the full evolution in Fig.\,\ref{fig:sim_gap} shows. Although this evolution is slow, it converges well within the lifetime of the disk. Therefore the simulations of \cite{Cazzoletti+17} and our simulations agree well with each other during the earlier stages, but our longer simulation time shows that the gap will widen with progressing evolution. We accordingly conclude that the observed gap size can be explained by the long-term action (10 000 orbits) of a binary with a separation of 35\,au that is coplanar with the disk.

We note that our choice for the viscous $\alpha$-parameter of $10^{-3}$ does not affect our conclusion. As the disk starts at more than 10\,au with rather low density, a low level of turbulence driven by the magnetorotational instability seems to be a reasonable assumption, and we consider our value a realistic choice. However, the relatively high mass of the disk may lead to an even lower viscosity. We therefore compared our results with a simulation using an even lower $\alpha$ parameter of $10^{-4}$. Our test run shows that lowering $\alpha$ affects the gap size only slightly, reducing it by less than 10$\%$.
The fact that in this case, a lower $\alpha$ viscosity slightly shrinks the gap size, is related to the relatively high binary eccentricity of GG\,Tau\,A ($\epsilon \sim$0.3). For eccentricities $\gtrsim$0.15, the eccentricity of the disk is directly affected by the binary eccentricity. Because the transfer of angular momentum is weaker, lowering $\alpha$ decreases the apocenter distance of the disk, while the pericenter distance remains constant, thus lowering the gap eccentricity and resulting in a slightly smaller net gap size (Penzlin et al. in prep.).
Similarly, \cite{Cazzoletti+17} observed no strong dependence of the location of the gas density peak on the assumed value of $\alpha$.
However, a lower $\alpha$ value would result in a significantly less eccentric gap.
Therefore a low viscosity may even be consistent with the fact that the disk does not appear very eccentric in the continuum observations.

In summary, our simulations suggest that a tight, $\sim$35\,au binary orbit that is coplanar with the outer disk is sufficient to create a gap in the disk of the observed size. However, we note that some misalignement within the system cannot be excluded, in particular, in view of the shadows on the outer disk, which may imply the presence of misaligned circumstellar material.
 Final conclusion on the orbital parameters of the binary and the respective disk-orbit orientation requires further astrometric monitoring as the current orbital coverage is still sketchy (Maire et al. in prep.).

%-----------------------------------------------------------------------
%-----------------------------------------------------------------------

\section{Summary and conclusions}\label{sec:conclusions}
We have observed the circumbinary environment of GG\,Tau\,A in polarized light with SPHERE/IRDIS in H band at unprecedented angular resolution. We analyzed the disk morphology and compared our observations to hydrodynamical simulations. The following section summarizes our findings.

The inner region appears to be highly structured. Our image suggests that the previously reported northern arc is composed of a double-arc structure. We furthermore detect various filament-like structures in the immediate circumbinary environment. Small dust grains scattering off light from the binary appear to be distributed in a large area around the binary.
We clearly detect previously suggested filament-like structures connecting the outer ring with the northern arc. The azimuthal spacing of the streamers may be consistent with a periodic perturbation by the binary, tearing off material from the inner edge of the outer disk once during each orbit. 
We confirm detection of three shadowed regions cast on the outer disk, as well as a tentative fourth shadow, suggesting the presence of an inclined circumstellar disk around Aa or Ab. We do not confirm a linear movement of the western shadow lane since 2011 that was suggested by previous observations.

We ran hydrodynamical simulations including the binary on an eccentric and disk coplanar orbit with a semimajor axis of 35\,au. The simulations ran for 28\,000 orbits, which covers the estimated age of the system. The final disk configuration shows evidence of spiral structures in the outer ring as well as within the cavity, similar to the observations.
 The resulting disk size is in qualitative agreement with the observations, which implies that a coplanar binary orbit $\sim$35\,au in size may be sufficient to explain the size of the ring. Astrometric follow-up observations are required to provide a final conclusion on the size and orientation of the binary orbit.

%-----------------------------------------------------------------------

\begin{acknowledgements}
       We thank the referee, Ruobing Dong, for providing constructive comments which helped improve the manuscript.
We thank A. M\"{u}ller, P. Pinilla, D. Price and J. Sanchez for fruitful discussions, as well as Y. Itoh for kindly sharing the previous Subaru data with us. M.K. warmly thanks Th. M\"{u}ller for help with the visualization of the data.
T.H. acknowledges support from the European Research Council under the Horizon 2020 Framework Program via the ERC Advanced Grant Origins 83 24 28.
MBe,FMe and MV acknowledge funding from ANR of France under contract number ANR-16-CE31-0013 (Planet Forming Disks).
C.P. acknowledges funding from the Australian Research Council via FT170100040 and DP180104235.
GL has received funding from the European Union's Horizon 2020 research and innovation programme under the Marie Sk\l odowska-Curie grant agreement No 823823 (RISE DUSTBUSTERS project).
A.Z. acknowledges support from the FONDECYT Iniciaci\'on en investigaci\'on project number 11190837.
SPHERE is an instrument designed and built by a consortium consisting of IPAG (Grenoble, France), MPIA (Heidelberg, Germany), LAM (Marseille, France), LESIA (Paris, France), Laboratoire Lagrange (Nice, France), INAF - Osservatorio di Padova (Italy), Observatoire de
Gen\`{e}ve (Switzerland), ETH Zurich (Switzerland), NOVA (Netherlands), ONERA (France), and ASTRON (The Netherlands) in collaboration with ESO. SPHERE was funded by ESO, with additional contributions from CNRS
(France), MPIA (Germany), INAF (Italy), FINES (Switzerland), and NOVA (The Netherlands). SPHERE also received funding from the European Commission Sixth and Seventh Framework Programmes as part of the Optical Infrared Coordination Network for Astronomy (OPTICON) under grant number RII3-Ct2004-001566 for FP6 (2004-2008), grant number 226604 for FP7 (2009-2012), and grant number 312430 for FP7 (2013-2016).
\end{acknowledgements}

%-----------------------------------------------------------------------

\bibliography{bibliography}
\bibliographystyle{aa}

%-----------------------------------------------------------------------

\appendix

\section{Polarized intensity pattern in the presence of two illumination sources}\label{sec:app:scattering}

\begin{figure*}[htp]
    \begin{centering}
        \includegraphics[width=\textwidth]{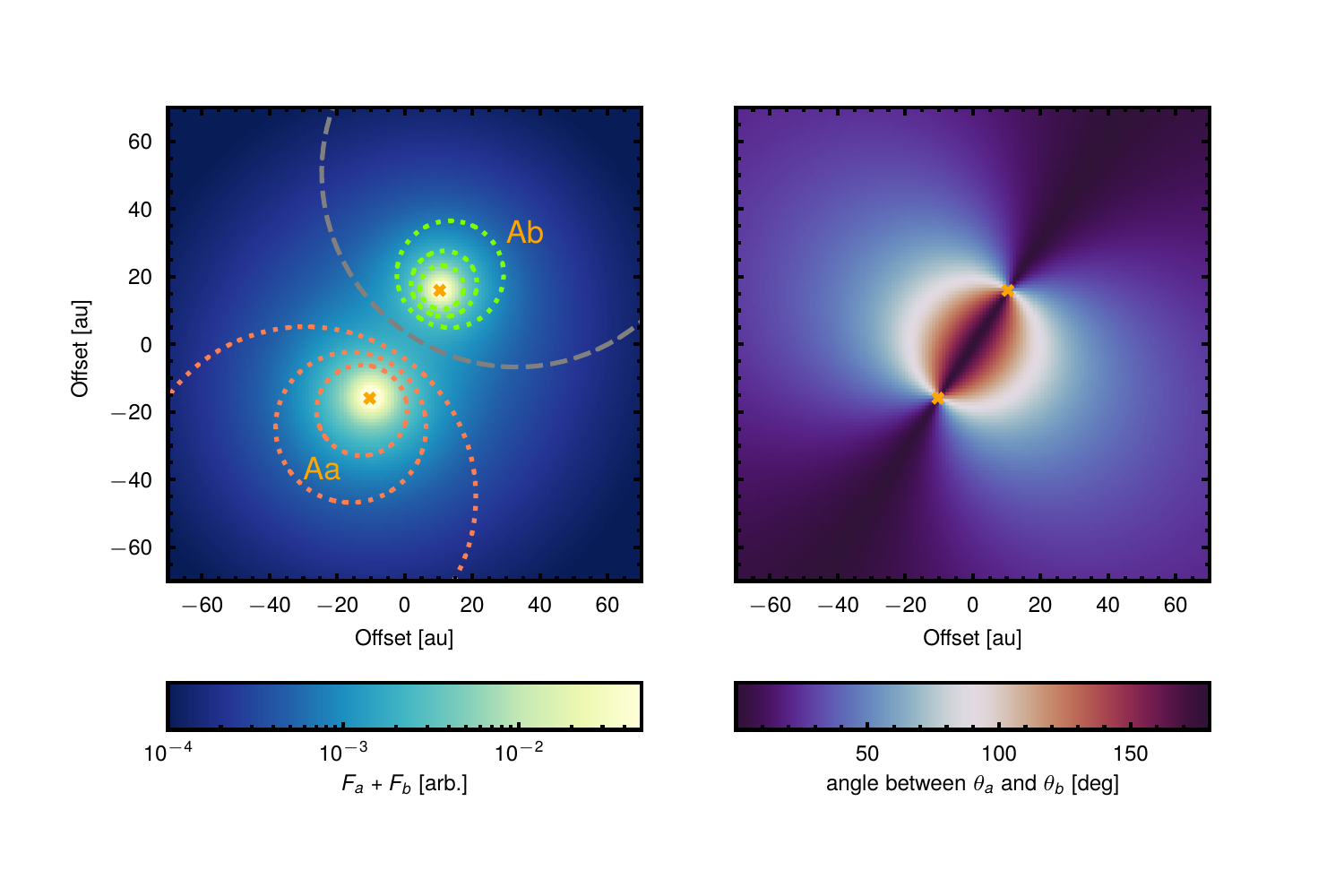}
    \end{centering}
    \caption{Toy model of the inner disk region of GG\,Tau\,A including two illuminating sources. The left panel shows the radiation field generated by the two stars. For each star, the dotted contours mark the regions where the contribution from the other star to the total flux is 5, 10, and 20\% (in red, the contribution from GG\,Tau\,Ab in the immediate surrounding of GG\,Tau\,Aa; in green, vice versa).
 The grey dashed line traces the contour where $F_a$ equals $F_b$. The right panel maps the angles between the linear polarization vectors resulting from scattering of light from Aa and Ab. In regions in which this angle becomes close to 90\degree, $PI$ could theoretically cancel out. }
    \label{fig:scattering}
\end{figure*}

In order to investigate how the presence of two illumination sources affects the morphology of \textit{PI}, we generated a toy model of the GG\,Tau\,A disk. We considered two illumination sources at a respective separation of 38\,au, and with luminosities ($L_a$=0.44\,\Lsun, $L_b$=0.23\,\Lsun) such as is found for GG\,Tau\,Aa and Ab. We assumed that the luminosity ratios of Aa and Ab are representative for their H-band flux ratios, which is consistent with the observations by \citet{diFolco+14}, who reported an H-band flux ratio of $\sim$2.1. For each point in the disk plane, we computed the received stellar illumination $F_{tot}=F_a+F_b\propto L_a/d_a^2+L_b/d_b^2$, where $d_a$ and $d_b$ is the distance to Aa and Ab, respectively. This received stellar flux is proportional to the intensity of scattered linearly polarized light, assuming a homogeneous surface density and degree of linear polarization throughout the disk. We also assumed a flat-disk geometry for simplicity. \\
Fig.\,\ref{fig:scattering} (left) shows the distribution of $F_{tot}$ for a face-on view of the disk. The dotted circles trace contours at which the contribution from the respective other star to $F_{tot}$ is 5, 10 and 20\%, that is, where $F_a/F_{tot}=(0.05,0.1,0.2)$ (green) and $F_b/F_{tot}=(0.05,0.1,0.2)$ (red). The 5\% contours are found as close as $\sim$9\,au ($\sim$5\,au) to the location of Aa (Ab). Because the PSF FWHM of our SPHERE observations is about 40\,mas (i.e., 6\,au at 150\,pc, thus corresponding to a PSF radius of $\sim$3\,au) the contribution of scattered polarized light from the respective other star to the unresolved polarized signal measured at the locations of GG\,Tau\,Aa and Ab is thus expected to be negligible. \\
We furthermore investigated whether any of our detected disk substructures might be related to the respective orientation of the polarization vectors in the presence of two illumination sources, rather than to a variation in disk surface density or scale height. If, for example, the polarization vectors at a certain point in the disk due to light scattered from Aa and Ab enclosed an angle of about 90\degree, the polarized signal could cancel out, leading to a locally depressed \textit{PI}. Because the orientation of the linear polarization vectors is expected to be orthogonal to the radius vectors connecting a certain point in the disk with the respective illumination sources, it is possible to map the angles enclosed by the two polarization vectors throughout the disk. This map is shown in Fig.\,\ref{fig:scattering} (right). A region with a respective polarization angle difference of 90\degree\ indeed lies close to the stars. Farther away, however, from $\sim$a$_{\mathrm{bin}}$ on, polarization vectors tend to be aligned with respect to each other. This is consistent with our observations, where the polarization vectors are clearly azimuthally orientated throughout the outer disk. While we cannot exclude that some spatial \textit{PI} variation close to the binary is caused by the superposition of the polarization vectors, we conclude that this effect cannot be responsible for the generation of any of the disk substructures we detected that are illustrated in Fig. \ref{fig:inner_region}.

%-----------------------------------------------------------------------
%-----------------------------------------------------------------------

\section{Early meta-stable simulation phase}\label{sec:app:stable}
\begin{figure}
    \begin{centering}
    \includegraphics[width=1\linewidth]{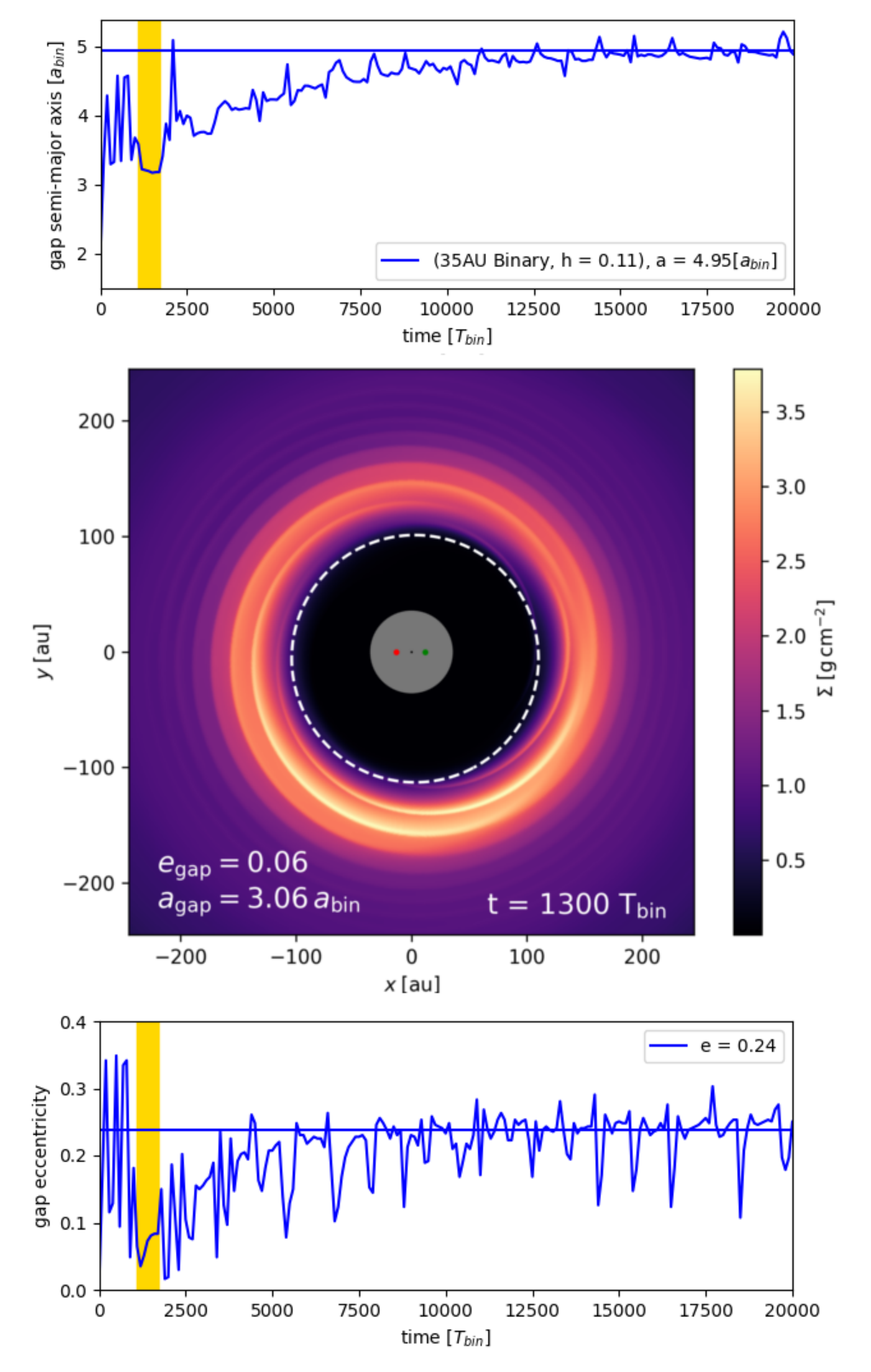}
    \end{centering}
   \caption{Gap size (top) and eccentricity (bottom) evolution of the cavity around GG\,Tau\,A created by our PLUTO simulation for the midplane temperature of the disk. The disk encounters a meta stable symmetric state (highlighted in yellow) with reduced gap size and eccentricity for about 600 orbits. The 2D surface density plot after 1300 orbits is shown in the middle. } 
\label{fig:cazo}
\end{figure}

\cite{Cazzoletti+17} found a smaller inner cavity after a simulation of about 1000 binary orbits. We can confirm this finding for the early simulation, as shown in Fig. \ref{fig:cazo}. However, we find that this feature is created by the symmetric initial condition of the gas distribution. After clearing the inner disk from gas in unstable orbits during the first few hundred orbits, the disk reaches a meta-stable configuration. This symmetric configuration will be disturbed by the higher modes of the binary potential and transform into the stable eccentric cavity that is reached after about $10\,000$ binary orbits. The same behavior occurs for less viscous systems a few hundred orbits earlier.

%\counterwithin{figure}{section}

\end{document}